\documentclass[revtex4]{emulateapj}

\def\feh{$[{\rm Fe/H}]$}
\def\teff{$T_{\rm eff}$}
\def\logg{$\log \, g$}

\def\ugriz{$ugriz$}
\def\recall{\textit{recall}}
\def\precision{\textit{precision}}

\def\pasa{PASA}

\usepackage[colorlinks,urlcolor=blue,citecolor=blue,linkcolor=blue]{hyperref}
\usepackage{booktabs}
\usepackage{natbib,amsmath,textcomp}
\slugcomment{DRAFT \today}
\shorttitle{Photometric Method to Discover EMP Stars}
\shortauthors{Miller}

\begin{document}

\title{The Synthetic-Oversampling Method: Using Photometric Colors to Discover Extremely Metal-Poor Stars}

\author{A.~A.~Miller\altaffilmark{1,2,3*}
}

\altaffiltext{1}{Jet Propulsion Laboratory, 4800 Oak Grove Drive, 
    MS 169-506, Pasadena, CA 91109, USA}
\altaffiltext{2}{California Institute of Technology, Pasadena, CA 91125, USA}
\altaffiltext{3}{Hubble Fellow}
\altaffiltext{*}{E-mail: {\tt amiller@astro.caltech.edu}}

\begin{abstract}

Extremely metal-poor (EMP) stars (\feh\ $\le -3.0$ dex) provide a unique window into understanding the first generation of stars and early chemical enrichment of the Universe. EMP stars are exceptionally rare, however, and the relatively small number of confirmed discoveries limits our ability to exploit these near-field probes of the first $\sim$500 Myr after the Big Bang. Here, a new method to photometrically estimate \feh\ from only broadband photometric colors is presented. I show that the method, which utilizes machine-learning algorithms and a training set of $\sim$170,000 stars with spectroscopically measured \feh, produces a typical scatter of $\sim$0.29 dex. This performance is similar to what is achievable via low-resolution spectroscopy, and outperforms other photometric techniques, while also being more general. I further show that a slight alteration to the model, wherein synthetic EMP stars are added to the training set, yields the robust identification of EMP candidates. In particular, this synthetic-oversampling method recovers $\sim$20\% of the EMP stars in the training set, at a precision of $\sim$0.05. Furthermore, $\sim$65\% of the false positives from the model are very metal-poor stars (\feh\ $\le -2.0$ dex). The synthetic-oversampling method is biased towards the discovery of warm ($\sim$F-type) stars, a consequence of the targeting bias from the SDSS/SEGUE survey. This EMP selection method represents a significant improvement over alternative broadband optical selection techniques. The models are applied to $>$12 million stars, with an expected yield of $\sim$600 new EMP stars, which promises to open new avenues for exploring the early universe.  

\end{abstract}


\section{Introduction}\label{intro}

Understanding the origins of structure on all scales, from the largest filaments containing galaxy clusters, to the smallest biological lifeforms that inhabit planets orbiting stars within the galaxies in those clusters, is arguably the main tenet of astronomy. The recent proliferation of wide-field surveys aims to study these problems, and a vast array of related questions, by generating large statistical samples that capture the diversity of different objects throughout the Universe. A challenge for these surveys, however, is that more data is not equivalent to better data. While the Large Synoptic Survey Telescope (LSST; \citealt{Ivezic08}) will eventually dwarf all other ground-based, wide-field optical surveys, the data deluge from LSST demands the development of superior algorithmic techniques. These methods must be capable of capturing and exploiting complex information from current and future data streams.

Data-driven methods, such as machine-learning algorithms, provide an intriguing solution to these challenges. These models are extremely flexible and have the ability to ascertain complex, non-linear interactions within the data. In brief, machine-learning models use a training set, a collection of sources with known \textit{labels}, such as a classification or physical property, to derive a mapping between those labels and \textit{features}, measured properties of the sources in the training set. Once the mapping is learned, this knowledge can be applied to new, unlabeled data. With spectroscopic resources already in short supply, a major challenge is deriving labels that are traditionally determined from spectroscopic measurements, e.g., redshift or metallicity, from photometric observations alone. The importance of solutions to this problem will be amplified during the LSST-era, when more than 20 billion sources will be photometrically detected \citep{Ivezic08}. The majority of these sources will not be amenable to spectroscopic observations, even with thirty-meter class telescopes.

Almost from the time it was realized that metal-rich stars produce less light in the blue optical than their metal-poor counterparts \citep{Schwarzschild55}, efforts have been made to photometrically estimate stellar metallicities (e.g., ultraviolet-excess technique; \citealt{Wallerstein62}). The most successful efforts to date use narrowband and mediumband filters, designed to be sensitive to metallicity dependent absorption lines in the stellar spectrum. The most prominent technique uses the $uvby\beta$ Str{\"o}mgren filters (see \citealt{Stromgren66} for a review), which have been demonstrated to produce \feh\footnote{Throughout this paper \feh\ is used as a proxy for metallicity, where \feh\ is defined as $\log (N_{\mathrm{Fe}}/N_{\mathrm{H}})_\ast - \log (N_\mathrm{Fe}/N_\mathrm{H})_\odot$, where $N_\mathrm{Fe}$ and $N_\mathrm{H} $ are the total number of iron and hydrogen atoms, respectively.} measurements with a scatter of $\sim$0.1 dex relative to spectroscopic observations for FG stars \citep{Nordstrom04}. For isolated groups of stars (clusters, galaxies), if there is a single stellar population (i.e., the were born during a single episode of star formation) and the distance is well known, then \feh\ estimates can be made using a single photometric color and isochrone fitting. Interestingly, \citet{Lianou11} find that isochrone fitting performs poorly relative to spectroscopic methods when the single stellar population assumption is violated. Modern wide-field surveys, such as the Sloan Digital Sky Survey (SDSS; \citealt{York00}) or LSST, primarily observe field stars with broadband filters. The age and distance to any field star is highly uncertain, meaning methods that use the SDSS filters are at a significant disadvantage relative to the Str{\"o}mgren filters or isochrone fitting. Nevertheless, when careful selections are made to limit samples to FG stars, broadband photometric estimates of \feh\ can be made with a typical scatter of $\sim$0.2--0.3 dex (e.g., \citealt{Ivezic08a,Bond10}). Achieving this precision requires the use of the $u$-band (see \citealt{An09}).

Stellar atmospheres retain the composition of the gas from which the star forms: as the universe becomes enriched with metals over time, so do newly formed stars. Thus, stellar metallicity measurements can serve as a proxy for stellar age (though the scatter in these relations is large, see \citealt{Soderblom10} for a review). Stars with very small metal abundances, known as extremely metal-poor (EMP) stars (\feh\ $\le -3.0$ dex), are relics from the early universe that provide unique insight to the nature of the first generation of stars. In particular, stars with $M_\ast/M_\odot \lesssim 0.8$ have not had sufficient time, within the age of the universe, to undergo significant post-main-sequence evolutionary changes and remain on, or close to, the main sequence. Therefore, the atmospheres of EMP stars retain information on the initial mass function of Population III stars, the diversity and nucleosynthetic yield of the first supernovae, the early chemical enrichment of the universe, and the formation of the first galaxies (for recent reviews on EMP stars see \citealt{Beers05,Frebel15}). As a result, considerable efforts have been made to identify EMP stars in the Milky Way halo in order to understand the nature of the Galaxy in the first $\sim$500 Myr after the Big Bang. 

Traditionally, candidate EMP stars are identified via objective-prism or low-resolution-spectroscopic surveys, and later confirmed via high-resolution spectroscopy. The HK Survey of \citet{Beers85,Beers92} identified EMP candidates from stars with weak \ion{Ca}{2} $K$ absorption. Several groups have utilized objective-prism observations from the Hamburg/ESO survey to identify EMP candidates and confirm bonafide EMP stars with high-resolution spectroscopy (e.g., \citealt{Cohen04,Frebel06,Christlieb08}). Recently, SDSS, and in particular the SDSS-II sub-survey known as the Sloan Extension for Galactic Understanding (SEGUE; \citealt{Yanny09}), have identified hundreds of EMP candidates from low-resolution spectra. Many of these candidates have been confirmed with high-resolution observations (e.g., \citealt{Aoki13}). Additional follow-up is on-going for all of these surveys, and more EMP discoveries can be expected. 

Early evidence of the utility of the ultraviolet-excess technique suggested that the relation saturated for very metal-poor (VMP) stars (\feh\ $\le -2.0$ dex), and this result has been seemingly confirmed with modern survey data (e.g., \citealt{Bond10}). As a result, there have been virtually no studies on the utility of identifying EMP stars from broadband photometric colors alone. Recently, \citet{Schlaufman14} developed a technique that exploits the significant near-infrared molecular absorption of metal-rich stars to identify candidate EMP stars. Using data from the Two Micron All Sky Survey (2MASS; \citealt{skrutskie-2mass}) and the \textit{Wide-field Infrared Survey Explorer} (\textit{WISE}; \citealt{Wright10}), \citeauthor{Schlaufman14} identify bright ($V < 14$ mag) EMP candidates, of which a small handful, corresponding to an efficiency of a few percent, have been confirmed via their initial follow-up spectroscopy. Additionally, the SkyMapper Telescope is poised to discover a large bounty of EMP stars by combining observations from the broadband $ugriz$ filters with a custom narrow filter centered on the \ion{Ca}{2}~K line \citep{Keller07}. The use of this narrow filter is extremely efficient for the discovery of EMP stars, and the early returns from SkyMapper include the confirmation of 41 EMP stars via high-resolution spectroscopy \citep{Jacobson15}. The unique filter combination has also led to the discovery of the most iron-poor star known \citep{Keller14}. SkyMapper follow-up is still ongoing, and estimates of the discovery efficiency using their narrow band filter are currently not available (though 41 of the 122 EMP candidates studied in \citealt{Jacobson15} were confirmed, suggesting an efficiency of $\sim$1/3). Nevertheless, this survey likely represents the premier method for uncovering southern sky EMP stars in the near future.

Here, a new technique to estimate \feh\ from only broadband $ugriz$ filters is presented. The method utilizes machine-learning algorithms and is trained using a sample of $\sim$170,000 stars with precise photometric observations and spectroscopic determinations of \feh\ from SDSS. It is demonstrated that the method is superior to other photometric \feh\ techniques. Furthermore, the method can be slightly altered, via the inclusion of synthetic EMP stars in the training set, to be suitable for the discovery of EMP stars. This final model enables the first-ever identification of EMP stars from broadband-optical filters alone.

\section{The Spectroscopic Sample}\label{sec:training_set}

Machine-learning models require a training set: a collection of sources with known labels. Once the mapping between features and labels is learned, the model can be applied to newly-observed, unlabeled sources for which only features are known. The construction of the training set and choice of machine-learning algorithm are essential steps for constructing a model that produces accurate predictions. Furthermore, as is the case for all data-driven approaches, the training set must be representative of the population of unlabeled sources or the model predictions will be unreliable. This is major challenge for many astronomical surveys: typically, new surveys probe fainter populations than those present in previously studied well-labeled samples (see e.g., \citealt{Richards12}). As detailed in \S\ref{sec:field_preds}, significant care is taken to ensure that the models developed here are only applied to the subset of field stars that is extremely similar to training set stars. 

\subsection{SDSS Spectroscopic Measurements of \feh}

SDSS is an optical, wide-field survey that has produced \ugriz\ imaging of $>  \,$14,500 deg$^2$ and collected spectra of $> \,$850,000 stars (several million spectra of extragalactic targets have also been obtained; \citealt{Alam15}). With $> \,$250,000,000 stars without spectroscopic observations, the SDSS dataset is ideal for the construction of the model: the large reservoir of spectroscopically observed stars will ensure a robust training set, yet there remains a significant pool of sources to search for candidate EMP stars. 

All SDSS optical stellar spectra are analyzed via the automated Segue Stellar Parameters Pipeline (SSPP; for full details on the SSPP see \citealt{lee08, lee08a, allende-prieto08}). Briefly, the SSPP determines \teff, \logg, and \feh\ for stellar sources using multiple parameter estimation methods (e.g., neural networks, synthetic spectral matching, \ion{Ca}{2} K line index technique, etc.). The individual measurements of \teff, \logg, and \feh\ are then robustly combined to provide final adopted values, and their corresponding uncertainties. For high signal-to-noise ratio (SNR) spectra with $4500 \, \mathrm{K} \le$ \teff\ $\le 7000 \, \mathrm{K}$, the SSPP determines \teff, \logg, and \feh\ with typical uncertainties of 157 K, 0.29 dex, and 0.24 dex, respectively. In addition to estimates of these parameters, the SSPP provides processing flags for sources where the parameter estimates are no good, such as white dwarfs or M stars. 

\subsection{Training Set Selection Criteria}

Photometric colors and spectroscopic \feh\ measurements for the training set sources are selected from SDSS data release 10 (DR10; \citealt{Ahn14}), which includes the most recent version of the SSPP. In total there are 427,225 sources with \feh\ measurements in DR10, however, this set is further pruned to avoid systematic biases and ensure a high-quality training set. The selection criteria are designed to select sources with the most reliable photometric and spectroscopic measurements. It is important to note, each of these criteria can be applied to the $\sim 2.6 \times 10^8$ SDSS stars with no spectroscopic observations, ensuring that these choices do not introduce a significant bias in the final model predictions. 

Poor, or missing, photometric measurements will corrupt the fidelity of the machine-learning models, thus, the first restrictions placed on the training set are photometric. The following photometric properties can all be retrieved from the \texttt{PhotoObjAll} table in the SDSS DR10 database. The first requirement for inclusion in the training set is a detection in each of the \ugriz\ bands, equivalent to \texttt{psfMag\_f} $> 0$, where \texttt{f} is the SDSS filter [427,177; in this and the next paragraph the number of sources remaining in the training set following each constraint will be given in brackets]. Good calibration in each filter, \texttt{calibstatus\_f} $=$ 1, where, again, \texttt{f} is the filter [420,575], and a non-flagged photometric measurement, i.e.\ \texttt{clean} $=$ 1 [399,646], are also required. Sources fainter than 19.5 mag in the $g$-band are excluded, \texttt{psfMag\_g} $\le 19.5$ [390,741]. Finally, sources with large photometric uncertainties are excluded, as these will result in a noisy mapping between colors and \feh. Sources with \texttt{psfMagErr\_u} $\ge 0.04$ mag, or \texttt{psfMagErr\_h} $\ge 0.03$ mag where \texttt{h} is any of the $griz$ filters are excluded [240,614].

Spectroscopic properties are retrieved from the DR10 \texttt{sppParams} table. The SSPP is most reliable for stars over a restricted range in \teff, $4500\, \mathrm{K} \le$ \texttt{TEFFADOP} $\le 7000\, \mathrm{K}$ \citep{lee08}, thus, stars outside this range are excluded [216,593].\footnote{Strictly speaking, \teff\ cannot be determined for stars with only photometric measurements. However, the photometric relation for \teff\ provided in \citet{Pinsonneault12} will enable the removal of stars that are too hot or too cool for the machine-learning model.} Furthermore, only stars with at least two individual measurements of \feh\ are included, \texttt{FEHADOPN} $\ge 2$ [209,163], as some of the individual SSPP methods for \feh\ measurements do not perform well over the full range of observed metallicities (see \citealt{Schlesinger12}). Requiring two \feh\ measurements significantly reduces the likelihood of a pathologically incorrect \feh\ measurement. Finally, only sources with the following SSPP flags are included: \texttt{nnnnn}, \texttt{nnngn}, or \texttt{nnnGn}, which correspond to normal stars, stars with a slight G-band feature, and stars with a potentially strong G-band feature, respectively [197,059]. Sources with any other combination of flags likely have unreliable \feh\ measurements and are unsuitable for this study (Y.\ S.\ Lee, private communication). Finally, for stars with multiple spectra only the highest SNR spectrum is retained in the training set [170,610].

These 170,610 stars form the training set, and the SSPP measured \feh\ values form the labels for the model. Prior to computing stellar colors, the observed brightness in each filter is de-reddened using the \citet{Schlafly11} recalibration of the \citet{Schlegel98}, hereafter SFD98, dust maps. 
The reddening corrected photometric colors, $(u - g)_0$, $(g - r)_0$, $(g - i)_0$, and $(g - z)_0$, constitute the full feature set for the model.

This reddening correction introduces some uncertainty into the model, however, the majority of SDSS observations are at high galactic latitudes, where extinction, and its corresponding correction, is small. The SFD98 dust maps measure the total Galactic reddening along a given sightline, meaning these corrections are equivalent to assuming the stars in this study reside outside the Milky Way. This assumption is clearly false, and maximally correcting for reddening in this way may result in some stars with colors that are too blue. Nevertheless, it is assumed that the bias from this overcorrection is small, especially because SDSS observations focused on low-extinction sight lines [$> 85$\% (92\%) of the training set has $A_r \le 0.2 \; (0.3)$ mag]. Furthermore, 
with a bright limit of $g \approx 14$ mag and a sample composed primarily of FG stars, the majority of stars in this study are $\gtrsim$1 kpc away, meaning the adopted reddening correction is reasonable. This assumption is further corroborated by the generally good performance of the model (see \S\ref{sec:results}). 

Reddening corrections become extremely problematic near the Galactic plane ($|b| \lesssim 10$\textdegree), where the SFD98 maps are unreliable and extinction is very patchy. As a result, the methods presented here will provide unreliable \feh\ estimates near the plane, unless superior extinction estimates to individual stars are developed. Further discussion of potential biases introduced by the reddening correction is provided in the conclusions (\S\ref{sec:conclusions}). 

\begin{figure*}
\centerline{\includegraphics[width=7.2in]{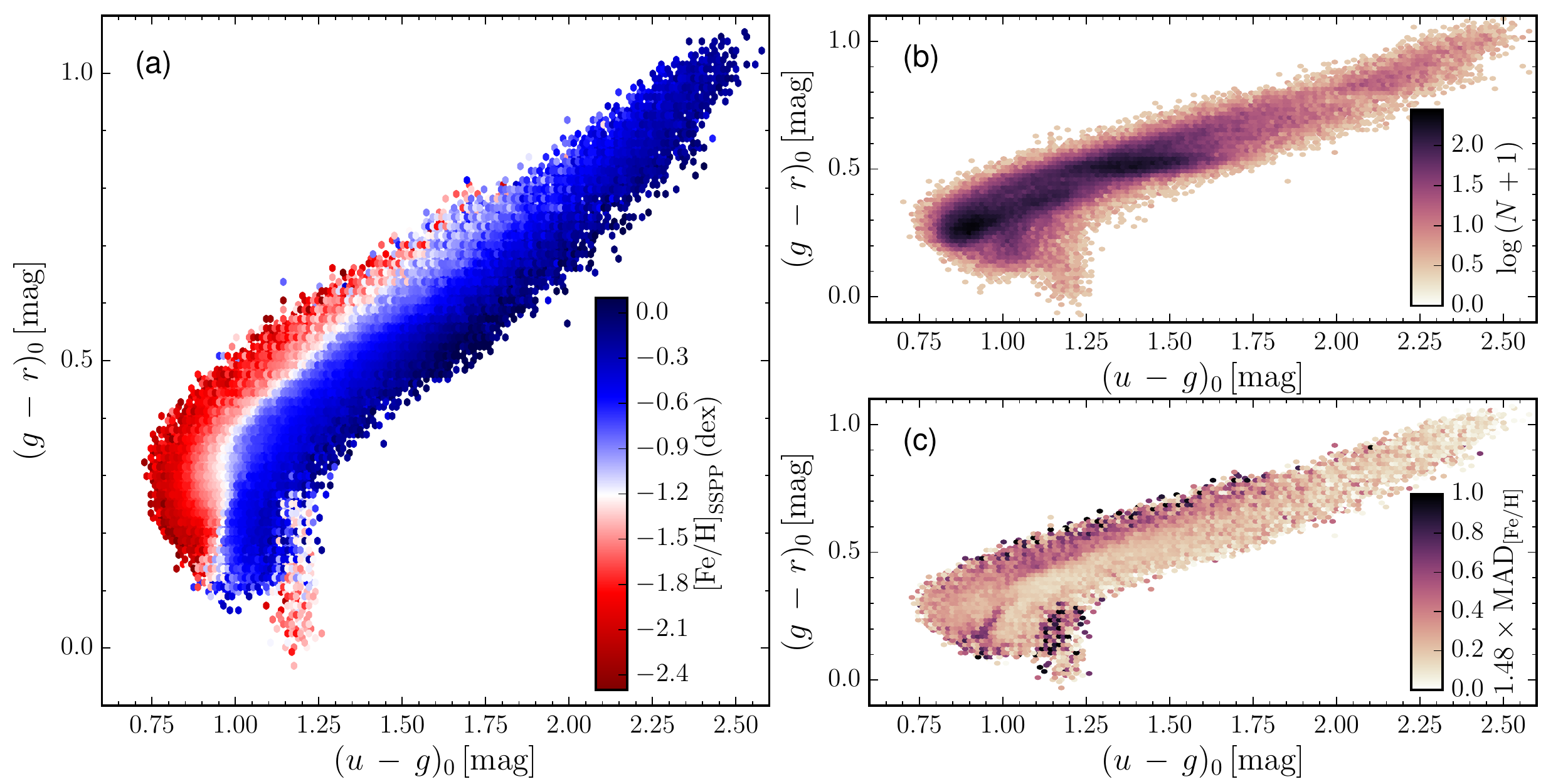}}
\caption[]{Summary of the training set shown in the $(u - g)_0$, $(g - r)_0$ CC diagram. Each plot shows summary statistics for the stars located within individual pixels, which are $\approx$0.01 mag on a side. Only pixels with $\ge$ 2 stars are shown. \textit{(a)}: the median \feh\ per pixel. Note that at constant $(g - r)_0$ color, roughly corresponding to constant \teff, the $(u - g)_0$ color provides an excellent diagnostic of \feh. \textit{(b)}: the total number of stars per pixel. Machine-learning methods typically perform best in regions where there is ample training data. The strong over-density with $0.48 < (g - r)_0 < 0.55$ is due to the SDSS emphasis on targeting G stars, while the over-density at $\approx (0.9, 0.3)$ is due to the targeting emphasis on F-turnoff-like stars (see e.g., \citealt{Yanny09}). \textit{(c)}: the scatter in \feh, as measured by 1.48 multiplied by the median absolute deviation (MAD), per pixel. Pixels with small scatter represent locations where the machine-learning model will be most accurate.}
\label{fig:SSPP_summary}
\end{figure*}

The scope of the training set is shown in Figure~\ref{fig:SSPP_summary}, which displays several summary statistics on a $u - g$, $g - r$ color-color (CC) diagram. The training set covers the full extent of the stellar locus, while spanning metallicities from EMP stars to metal-rich stars (\feh\ $> 0.0$ dex). Figure~\ref{fig:SSPP_summary}(a) demonstrates that \feh\ can be readily determined from photometric colors. 

\section{Machine Learning Models}\label{sec:models}

In addition to building a robust and representative training set, the choice of machine-learning algorithm is essential for the construction of a useful model.\footnote{For a general overview of machine learning, we refer the interested reader to \citet{Hastie09}.} Below a brief overview of the three different algorithms utilized in this study is provided. 

\subsection{$K$-nearest neighbors} 

$K$-nearest neighbors (KNN) regression identifies the $K$ training-set sources that are closest to the newly observed source in feature space. For the new source, the predicted value of the target variable is simply the mean from the $K$ neighbors. An advantage of KNN regression is that it is simple, and the model results are easy to interpret. In this study, KNN regression is performed using the \texttt{Python scikit-learn} implementation of the algorithm \citep{Pedregosa11}. Scaling factors are applied to each individual color so that the re-scaled features have a sample mean of zero and sample variance unity prior to performing KNN regression. 

\subsection{Random forest}

Random forest (RF) methods utilize the aggregation of multiple decision trees to assign a final classification or regression value to newly observed sources \citep{Breiman01}. RF makes use of bagging (see \citealt{Breiman96}), wherein bootstrap samples of the training set are used to construct each of the $N_{\rm tree}$ total trees in the forest. As each tree in the forest is constructed, only a random subset of $m_{\rm try}$ features is selected from the full feature set as a potential splitting criterion at each node of the tree. The use of bagging and $m_{\rm try}$ random features reduces the variance of the final model predictions relative to single decision-tree models, providing low-bias, low-variance results. The final RF predictions are determined by averaging the predictions for a new source from each of the $N_{\rm tree}$ individual trees. Furthermore, the RF algorithm is fast, each of the trees can be constructed independently and thus in parallel, and relatively easy to interpret. RF models have recently become highly popular as an application for astronomical data sets due to their relative insensitivity to noisy or meaningless features (e.g., \citealt{Brink13, Miller15}), and their invariant response to even highly non-gaussian feature distributions (e.g., \citealt{Dubath11,Richards11}). This study utilizes the \texttt{Python scikit-learn} implementation of the RF algorithm \citep{Pedregosa11}.

\subsection{Support vector machines}

Support vector machines (SVMs; \citealt{Boser92,Cortes95}) are learning models that project the features from the training set into a high- or infinite-dimension space. SVMs then find a linear hyperplane with the maximal margin separating the two groups of sources, in the case of classification. These methods can be generalized to regression problems \citep{Drucker97}, where the hyperplane must produce predictions on the training set that are within a given threshold of their true values. For this study a non-linear radial basis function is used to perform SVM regression, which is implemented using the \texttt{LIBSVM} software package \citep{Chang11}. For the SVM model, the individual colors are re-scaled so that the minimum and maximum values of the features are 0 and 1, respectively. 

\section{Regression Model Results}\label{sec:results}

\subsection{Comparison of the Three Regression Models}\label{sec:regress_results}

\begin{figure*}
\centerline{\includegraphics[width=7.4in]{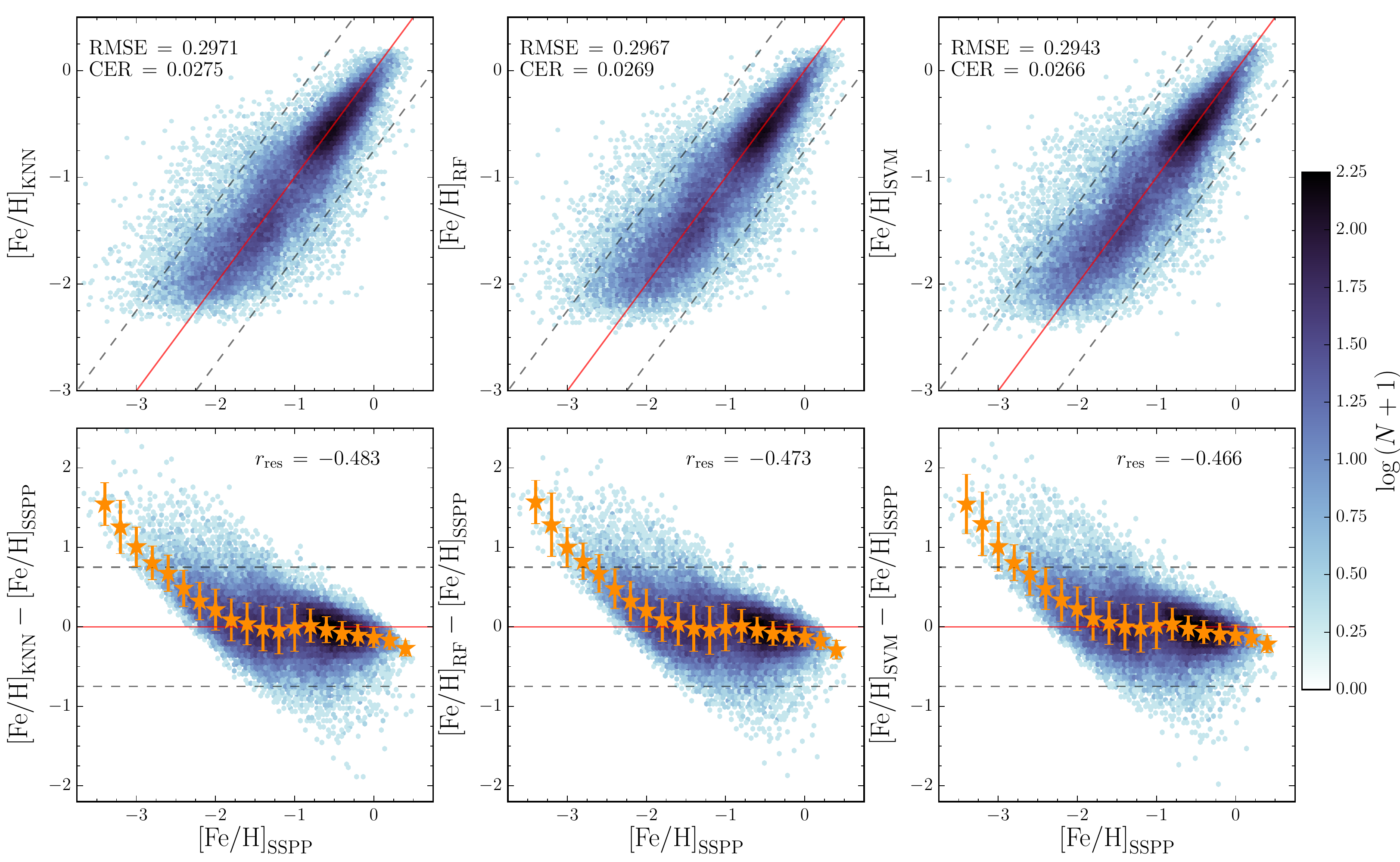}}
\caption[]{Regression results for the optimized KNN, RF, and SVM models shown, respectively, in the three columns from left to right. \textit{Top}: Density plot showing the number of sources in each pixel on the predicted \feh\ vs.\ SSPP \feh\ plane. Pixels are $\sim$0.05 dex on a side. The solid, red line shows the relation for perfect 1:1 regression, while the dashed, grey lines show the boundaries for catastrophic errors, $\pm0.75$ dex. Sources located outside the grey lines are considered catastrophic outliers. The SVM model has the smallest RMSE and CER. \textit{Bottom}: Residuals from the models (shown above), with the density of sources shown in each pixel. The orange stars show the median residual value in bins of width 0.2 dex. The associated errorbars show the scatter in each bin: 1.48$\times$MAD, an outlier-resistant and robust measure of the scatter. $r_{\rm res}$ is the Pearson $r$ correlation coefficient for the residuals as a function of \feh$_{\rm SSPP}$. $r_{\rm res}$ values close to zero indicate little bias in the model predictions. The SVM model produces the least biased estimates of \feh.}
\label{fig:regress_residuals}
\end{figure*}

To determine which of the three models from \S\ref{sec:models} best generalizes to new data, the 170,610 spectroscopic sources were separated into a training set containing 110,000 sources and a validation set with 60,610 sources. The models are optimized via a grid search over the relevant tuning parameters using 10-fold cross validation (CV) performed on the 110,000 source training set. The parameters that minimize the root-mean-square error (RMSE):
$${\rm RMSE} = \left[ \frac{1}{n}\sum_i^n(y_i - x_i)^2\right]^{1/2},$$ 
where $n$ is the total number of sources in the training set, $y_i$ is the model prediction of \feh\ for the $i^{th}$ source, and $x_i$ is the \feh\ spectroscopic value for the $i^{th}$ source. Small changes in the optimal tuning parameters do not significantly alter the CV RMSE. The optimal models were applied to the 60,610 source validation set, with the results summarized in Table~\ref{tbl:final_regress} and the final predictions shown in Figure~\ref{fig:regress_residuals}.

\begin{deluxetable}{lrrr}
\tabletypesize{\small}
\tablecolumns{4}
\tablewidth{240pt}
\setlength{\tabcolsep}{6pt}
\tablecaption{Test Set Predictions\label{tbl:final_regress}}
\tablehead{\colhead{Model} & \colhead{RMSE} & 
    \colhead{CER} & \colhead{Train Time\tablenotemark{a}} \\
    \colhead{} & \colhead{(dex)} & \colhead{} & \colhead{(s)}}
\startdata
KNN & 0.297 & 0.028 &    0.1 \\
RF & 0.297 & 0.027 &   72.4 \\
SVM & 0.294 & 0.027 &  728.4
\enddata
\tablecomments{Models: KNN -- k-nearest neighbors, 
        RF -- random forest, SVM -- support vector machines. All models have 
        been optimized using 10-fold cross validation and a grid search of 
        their respective tuning parameters. The results shown here reflect 
        the average of 5 separate models, and the model-to-model scatter is 
        small.}
\tablenotetext{a}{Average time required to train the 
        model on 110,000 sources using dual-socket, 8-core, 2.66 GHz Intel 
        Sandy Bridge CPUs with 64 GB of memory.}
\end{deluxetable}

The panels in Figure~\ref{fig:regress_residuals} show that the KNN, RF, and SVM models all produce similar predictions for \feh\ based on photometric colors. Formally, the SVM model produces the best predictions with RMSE $= 0.2943$ dex, which is $\sim$1\% better than the KNN and RF models. The SVM model also has the lowest catastrophic error rate, CER, defined as the fraction of sources where the predicted and spectroscopic values of \feh\ differ by $\ge 0.75$ dex. Again, while the SVM model has the best performance the difference between the three is small $\sim$1--3\%. The residuals, shown in the bottom panel of Figure~\ref{fig:regress_residuals}, are also similar for the three models. For stars with \feh$_\mathrm{SSPP}$ between $\sim$0 and $-2$, corresponding to the vast majority of stars in the Galaxy \citep{Schlesinger12}, the models exhibit virtually unbiased predictions with small scatter. There is a systematic bias for stars with \feh$_\mathrm{SSPP}$ $\lesssim -2$ or \feh$_\mathrm{SSPP}$ $\gtrsim 0$, which have over- and under-predicted values of \feh, respectively.

\subsection{Understanding the Regression-Model Bias}

As a measure of the overall bias of each model, the Pearson $r$ correlation coefficient is measured for the residuals as a function of spectroscopic \feh\ values. An unbiased model would show little to no correlation, $|r| \approx 0$. Models with $|r| \rightarrow 1$ show a strong correlation between the residuals and \feh, indicating significant bias in the final model predictions. The SVM model has the smallest $|r|$, meaning it has the smallest bias of the three machine-learning models. 

The correlation between the residuals and either \teff, the individual photometric colors, or \logg, is significantly weaker than the correlation between the residuals and \feh$_{\rm SSPP}$. Using the Fisher transformation of the Pearson $r$ coefficient, the correlation of the residuals with each of the photometric colors, \teff, and \logg\ is significantly smaller, probability $P \ll 0.0001$, than the correlation with \feh.

Thus, the systematic biases seen in Figure~\ref{fig:regress_residuals} are most likely the result of alternative effects. There are two systematic effects that play a role in this bias: (i) regression to the (sample) mean, and (ii) regression dilution bias. Non-parametric, data-driven regression models often produce predictions biased towards the sample mean. This effect can most easily be illustrated for KNN models. Consider the most metal-poor star in the validation set, which has \feh\ $= -3.68$ dex, at best, the KNN prediction for this source would be the mean \feh\ of the 60 most metal-poor stars\footnote{$K = 60$ for the optimized KNN model.} from the training set, which is equal to -3.38 dex. This represents the best case scenario, if the nearest neighbors for this EMP star include any that are not the least metal-poor in the training set the model-predicted \feh\ will be biased even further from the true value. The models are also susceptible to bias due to the uncertainties associated with the photometric colors and spectroscopic \feh\ measurements. Noisy features and target variables lead to a flattening of the regression slope, an effect known as regression dilution bias \citep{Frost00}. This bias could be improved in the future with more precise color measurements and superior spectroscopic determinations of \feh, though it may be prohibitively expensive to obtain these observations. Further discussion of these two types of bias can be found in \citet{Miller15}.

Physical effects may also be responsible for the systematic over-prediction of \feh\ for VMP stars. As metals are removed from a stellar atmosphere, the absorption lines present become weaker and weaker. Eventually, at some critical metallicity, $Z_\mathrm{crit}$, the lines will become so weak that they can no longer be detected via broadband-photometric colors. This means photometric-metallicity techniques eventually saturate, and assign all stars with $Z < Z_\mathrm{crit}$ the same \feh. If $Z_\mathrm{crit}$ occurs at \feh\ $\approx -2.0$ dex, then this would naturally explain some of the bias seen in Figure~\ref{fig:regress_residuals}. The photometric technique presented in \citet{Bond10} shows a similar saturation for stars with \feh\ $\lesssim -2.0$ dex. Nevertheless, in \S\ref{sec:synthetic_oversampling} it is shown that EMP stars can be recovered using broadband optical colors, meaning the saturation of photometric metallicity is not solely responsible for the biased VMP star predictions. 

\subsection{Comparison to spectra}

With an RMSE scatter of $\sim$0.29 dex, the SVM model produces predictions of \feh\ that are similar to those from low-resolution spectra. The SSPP provides \feh\ measurements with a typical uncertainty of $\sim$0.24 dex \citep{lee08}, though this precision is limited to stars with high SNR ($\gtrsim$ 50). To better facilitate the comparison between this study and the SSPP results, we compare the scatter between the 77 stars in this study that are also part of the SSPP high-resolution validation set (see Tables~3~and~4 of \citealt{allende-prieto08}).\footnote{Most of these 77 stars are in the 110,000 star training set. Thus, the SVM predictions here are from 10-fold CV to avoid an overlap between the training and test sets.} Adopting the \feh\ values measured from the high-resolution spectra as ground-truth, then the SSPP has an RMSE $= 0.37$ dex, while the SVM model has an RMSE $= 0.44$. With the caveat that this comparison is based on a small number of stars, this suggests that the SSPP performs $\sim$17\% better than the SVM model. These RMSE values are significantly larger than those reported in \citet{lee08}, because the 77 stars in common between this study and the sample in \citet{allende-prieto08} primarily excludes relatively metal-rich stars. The analysis presented in \citet{allende-prieto08} consists of two samples: a relatively metal-rich sample (median \feh\ $\approx -0.5$ dex) observed with the Hobby Eberly Telescope (HET), and a relatively metal-poor sample (median \feh\ $\approx -2.0$ dex) observed with the Keck and Subaru telescopes. The initial study compares 81 stars observed with HET and 44 stars observed by Keck and Subaru, while the sample in common with this study retains only 42 HET stars and 35 from Keck and Subaru. The scatter between the SSPP and the high-resolution measurements from Keck and Subaru (0.41 dex; see Table~6 of \citealt{allende-prieto08}) is significantly worse than the scatter for stars observed with HET (0.12 dex). Both the SSPP and the SVM model perform better on relatively metal-rich stars, thus, the preferential exclusion of these stars in the HET sample would lead to a corresponding increase in the RMSE. 

\subsection{Comparison to other photometric methods}

With photometrically observed stars in SDSS outnumbering spectroscopically observed stars by nearly a factor of $\sim$10$^3$, there have been many efforts focused on determining photometric metallicity estimates from broadband SDSS colors. In \citet{Kerekes13}, a KNN method is used to predict \feh\ with an RMSE $\approx 0.32$ dex for stars with $15 \, {\rm mag} < g < 17 \, {\rm mag}$, and 0.41 dex for stars with $18 \, {\rm mag} < g < 19 \, {\rm mag}$. The sample in the \citeauthor{Kerekes13}~study places no restrictions on the quality of the photometric or spectroscopic observations. Thus, stars that raised SSPP flags or have large photometric uncertainties are likely driving the significantly larger RMSE from that model. 

Multi-dimensional polynomial fits to the median \feh\ in 0.02 mag$^2$ bins in the $(u - g)_0$, $(g - r)_0$ plane are used to determine photometric metallicities in \citet{Ivezic08a}. This method is later updated in \citet{Bond10}, where SSPP values from DR7 replace the less accurate values from DR6, which were used in the \citeauthor{Ivezic08a} study. The fit presented in \citet{Bond10} produces a typical RMSE $\sim$0.2 dex for metal-rich stars and $\sim$0.3 dex for metal-poor stars. These values cannot be directly compared to those presented in \S\ref{sec:regress_results}, however, as the samples used in both the \citeauthor{Ivezic08a}\ and \citeauthor{Bond10}\ studies placed more stringent cuts on the training set than those employed here. In particular, those studies included only sources with $0.2 < (g - r)_0 < 0.6$, so as to focus on F/G stars. If the same selection criteria from \citet{Ivezic08a} are applied to the validation set from this study, 35,377 of the 60,610 stars remain. The RMSE for those stars is $\sim$0.26 dex for the SVM model and $\sim$0.32 dex for the photometric model presented in \citet{Bond10}.\footnote{See their Equation~A1, which is an update to Equation~4 presented in \citealt{Ivezic08a}.} Thus, the SVM model presented in this study represents an $\sim$18\% improvement in the scatter relative to the polynomial-fitting method presented in \citet{Ivezic08a} and \citet{Bond10}. 

\section{Model Alterations to Emphasize the Selection of EMP Stars}\label{sec:synthetic_oversampling}

While the regression models presented in \S\ref{sec:results} perform well for the vast majority of field stars, the strong biases for VMP stars make it  difficult to identify EMP stars. The discovery of EMP stars can be cast as a classification problem where all EMP stars belong to one class and all other stars, with \feh\ $> -3.0$, form the other class. For the 170,610 stars in this study, 256 are EMP stars. Thus, there is a significant class imbalance between the EMP and non-EMP stars. Typically, machine-learning classification algorithms are built to maximize the overall accuracy of predictions. A classifier that predicts all stars belong to the majority non-EMP class would have an accuracy of 99.8\%. For most machine-learning models this accuracy would be stunning. This masks the failure of the model for its most interesting task: identifying new EMP stars. Following some adjustments to the training set, however, EMP stars can be reliably recovered.

\subsection{Dealing with Class Imbalance: Upsampling and Downsampling}

Many classification problems deal with imbalance, wherein at least one class represents a very small minority of sources. It is often the case, however, that the minority class represents the target of interest: identifying additional instances of these rare events is the motivation for model construction. Minimizing the overall classification error rate means special attention is not paid to the minority class and these sources are disproportionally misclassified. The consequences range from mildly annoying, e.g., spam email bypassing filters to reach an inbox, to extremely serious, e.g., in the medical profession.

There are two general approaches for dealing with class imbalance. One approach is to manually adjust the imbalance in the training set by randomly downsampling the majority class or oversampling the minority class, or using a combination of the two. The other is to use cost-sensitive learning, where the cost for misclassification of the minority class is higher than the cost for misclassifying members of the majority. Most efforts focus on some form of the sampling technique, with downsampling approaches typically outperforming oversampling (see e.g., \citealt{Chen04}). A downside to downsampling is that information is being removed from the classifier, while strict oversampling will always be fundamentally limited by the fact that no truly new instances of the minority class have been added to the classifier.

Many researchers have found that over-sampling the minority class with replacement does not significantly improve minority-class recognition (e.g., \citealt{Ling98}). As a result, \citet{Chawla02} developed the synthetic minority over-sampling technique (SMOTE), wherein synthetic members of the minority class are generated to reduce the class imbalance. In short, synthetic members are generated by fitting a KNN model to the minority class. For each minority-class source in the training set, one of the $k$-nearest neighbors is selected at random, and a synthetic member of the minority class is generated by selecting a random point along the feature vector connecting the source and its neighbor. This process is then repeated to achieve the desired amount of oversampling. While examining a variety of classic class-imbalance problems, SMOTE outperforms over-sampling, while the combination of SMOTE and downsampling performs better than both downsampling and cost-sensitive learning methods \citep{Chawla02,Chawla03}. 

In \citet{Chen04}, two methods, which leverage both the sampling and cost-based approaches, are explored to improve the performance of RF on imbalanced data. The first approach, which they refer to as balanced random forest, uses a bootstrap sample of the minority class as well as an equal number of majority class members selected randomly with replacement to initiate each tree in the forest. Thus, the minority and majority classes are equally balanced over the classifier. The other method, which they refer to as weighted random forest, places a stronger penalty on misclassifying the minority class by weighting the samples when selecting splitting criteria at each node within individual trees and also weighting the final vote in the terminal nodes of each tree. Using multiple different performance measures, both methods show improvements relative to other techniques, including SMOTE plus down-sampling, over a variety of different problems \citep{Chen04}. 

\subsection{Improving Minority Class Recognition with Synthetic EMP Stars}\label{sec:sym_EMP}

Initial tests of the three methods described above, SMOTE, balanced RF, and weighted RF, showed no significant improvement in the recovery of EMP stars. This is likely the case due to the extreme imbalance for the problem at hand: the minority class constitutes less than 0.2\% of the training set, which is significantly less than the datasets tested in \citet{Chawla02,Chawla03} and \citet{Chen04}. Instead, a SMOTE-inspired approach, which generates synthetic EMP stars in a different manner, is developed.

\begin{figure}[ht]
\centerline{\includegraphics[width=3.6in]{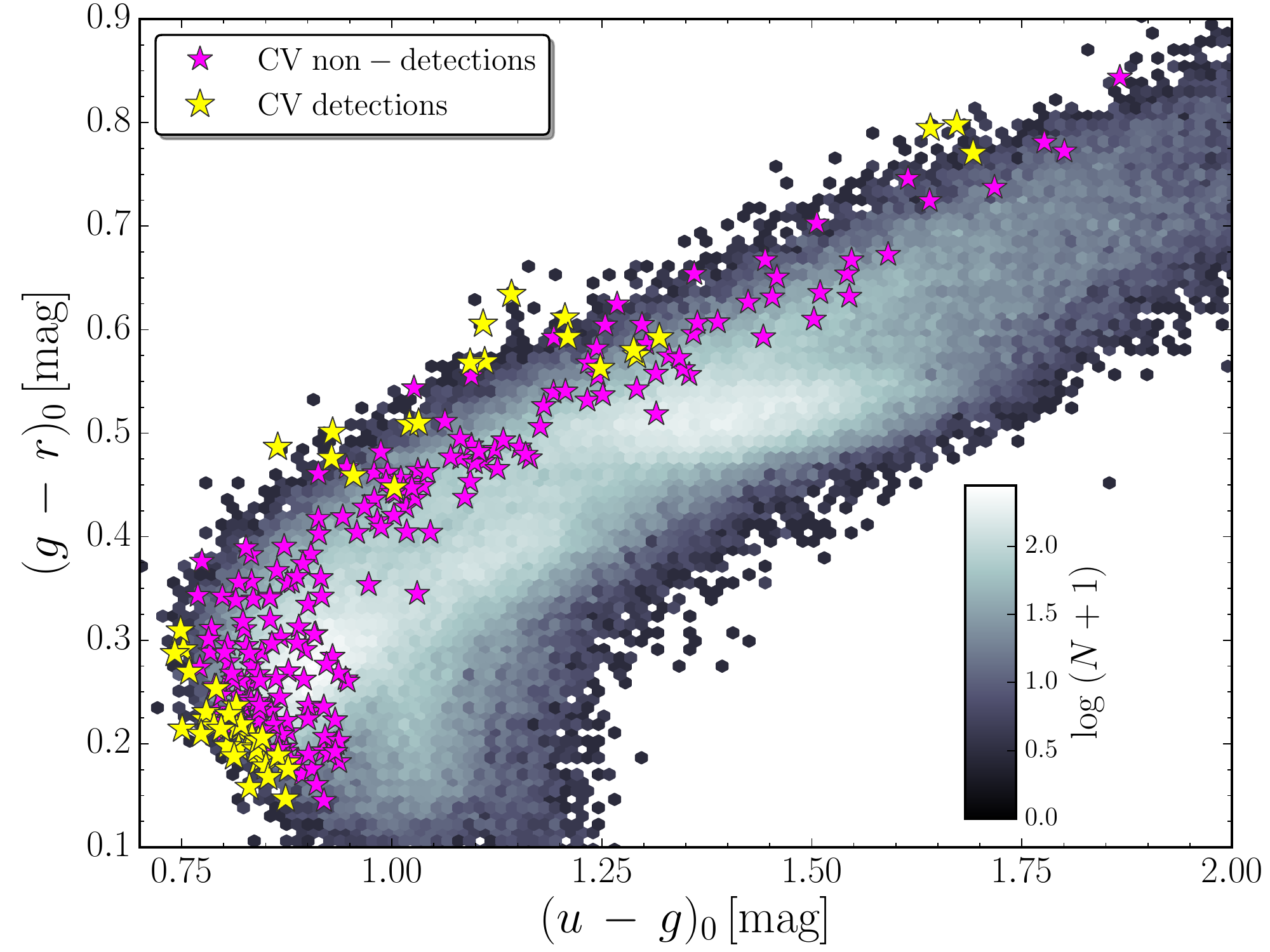}}
\caption[]{$(u - g)_0, (g - r)_0$ CC diagram showing the density of sources in the training set. The location of EMP stars in the training set is highlighted. Notice that the EMP stars form a relatively tight cluster that is parallel to the main stellar locus. EMP stars that are detected in CV via the synthetic-oversampling method are shown in yellow, while non-detections are shown in magenta. The synthetic-oversampling method is biased towards recovering those stars on the extremes of the sample distribution (see also \S\ref{sec:EMP_bias}).}
\label{fig:EMP_locations}
\end{figure}

A zoom-in on the $(u - g)_0$, $(g - r)_0$ CC diagram is shown in Figure~\ref{fig:EMP_locations} with the location of the EMP stars in the training set highlighted. While there is a relatively tight cluster of EMP stars on the blue edge of the stellar locus, approximately centered at (0.8, 0.2), this roughly coincides with the highest density location of non-EMP stars. Furthermore, over half the EMP stars form a loose sequence along the upper portion of the stellar locus. Thus, SMOTE, which generates synthetic samples between nearest neighbors while ignoring any underlying structure, is liable to create synthetic EMP stars that lie off the relatively well defined sequence. Instead, a different approach, which I refer to as the synthetic-oversampling method, is adopted: new EMP stars are generated by resampling the photometric colors within the reported uncertainties from SDSS. In practice, the procedure is straightforward: EMP stars are selected randomly with replacement from the training set. The photometric measurements for each of the $ugriz$ filters are then adjusted via a random number selected from a normal distribution with mean zero and standard deviation equal to the SDSS-measured photometric uncertainty in the respective filter. Colors for the synthetic stars are computed, and the SSPP \feh\ measurement for the original star is assigned to the synthetic star. Finally, the user may specify how many synthetic EMP stars are generated and included in the training set. 

As previously noted, the overall accuracy of a classifier is a poor measure of performance when trying to identify minority-class members in extremely imbalanced problems. Instead of focusing on the true positive rate and false positive rate, respectively, I aim to simultaneously maximize the \textit{precision} and \textit{recall} of the model, which are defined as: 
\begin{equation}
    \label{eq:precision_recall}
    \begin{split}
    precision = \frac{\textrm{TP}}{\textrm{TP} + \textrm{FP}}, \\
    recall = \frac{\textrm{TP}}{\textrm{TP} + \textrm{FN}},
    \end{split}
\end{equation}
where TP is the number of true positives, FP is the number of false positives, and FN is the number of false negatives.\footnote{Note that there are many different nomenclatures throughout the machine-learning literature for the terms defined in Eqn.~\ref{eq:precision_recall}. \recall\ is most commonly referred to as the true positive rate (TPR), though it can also be referred to as the sensitivity, hit rate, or completeness depending on the context. I adopt the convention of referring to this as the \recall\ as this is only discussed relative to the \precision. The \precision\ of a model is sometimes referred to as the positive predictive value or purity.} The \precision\ is a measure of how many erroneous measurements are required to recover a new member of the minority class, while the \recall\ is a measure of the fraction of the minority class that is actually recovered. Ideally, a model would produce a \precision\ and \recall~$\approx 1$, however, in practice, one must adopt a candidate decision threshold that offers a trade off between these two desirable features. 

The SVM model is adopted to search for EMP stars due to its superior performance in the \feh\ regression problem discussed in \S\ref{sec:results}. An SVM regression model is used, instead of an SVM classification model, so that candidates may be ranked by their likelihood of belonging to the EMP class. Thus, unlike a classification model where a single hard boundary between classes is determined, the class boundary from the regression model can be varied across different values of \feh\ to determine the optimal trade off between \precision\ and \recall. Given the rarity of EMP stars, the model is optimized via cross validation over the entire 170,610 source training set, rather than splitting the data into a training and validation set as was done in \S\ref{sec:results}. The SVM tuning parameters are optimized via three different instances of 10-fold cross validation to maximize the \recall\ at a \precision\ of 0.05, which is adopted as the figure of merit (FoM). This FoM corresponds to only one in every 20 EMP candidates identified by the model being a genuine EMP star. While this performance seems relatively poor, it represents a dramatic improvement over previous broadband photometric techniques to identify EMP stars. 

The results from optimized models with differing amounts of downsampling and synthetic oversampling are summarized in Table~\ref{tbl:EMP_preds}. As a baseline for the increase in performance downsampling and synthetic oversampling provide, the results for the full training set, with no synthetic over-sampling or downsampling, are also included. In addition to the FoM, Table~\ref{tbl:EMP_preds} also includes other measures of model performance for imbalanced problems, including: the receiver operating characteristic (ROC) area under the curve (AUC), the \precision-\recall\ AUC, and the $F$-measure, defined as: 
$$F\textrm{-measure} = \frac{2 \times precision \times recall}{precision + recall}.$$
For Table~\ref{tbl:EMP_preds} the $F$-measure is determined using a classification boundary of \feh$_{\textrm{SVM}} = -3.0$ dex, where sources with a predicted \feh\ below this value are considered EMP candidates. The ROC curve traces the trade off between the true positive rate and the true negative rate as a function of classification decision boundaries. The ROC AUC measures the area beneath the ROC curve and is used to evaluate the overall performance of a classification model. The closer the ROC AUC is to 1, the better the model, though note that this metric does not consider \precision. The \precision-\recall\ AUC measures the overall performance of a classifier on imbalanced data. Again, the closer this value is to 1, the better the model.

\begin{deluxetable*}{crcccccc}
\tabletypesize{\small}
\tablecolumns{8}
\tablewidth{0pt}
\setlength{\tabcolsep}{6pt}
\tablecaption{Optimized EMP classification results\label{tbl:EMP_preds}}
\tablehead{\colhead{ds\tablenotemark{a}} & 
    \colhead{$N$\tablenotemark{b}} & 
    \colhead{ROC AUC} & \colhead{PR AUC} & 
    \colhead{$F$-measure} & 
    \colhead{$R(P = 0.1)$\tablenotemark{c}} & 
    \colhead{$R(P = 0.05)$\tablenotemark{d}} & 
    \colhead{$P(R = 0.1)$\tablenotemark{e}} }
\startdata
 25 &   0 & 0.909 & 0.022 & 0.000 & 0.000 & 0.132 & 0.049  \\
 25 & 1000 & 0.916 & 0.027 & 0.064 & 0.022 & 0.197 & 0.052  \\
 25 & 2000 & 0.920 & 0.026 & 0.078 & 0.016 & 0.198 & 0.056  \\
 25 & 3000 & 0.921 & 0.025 & 0.074 & 0.007 & 0.193 & 0.052  \\
 25 & 4000 & {\bf 0.922} & 0.025 & 0.066 & 0.003 & 0.178 & 0.053  \\
 50 &   0 & 0.909 & 0.023 & 0.000 & 0.001 & 0.133 & 0.055  \\
 50 & 1000 & 0.915 & 0.027 & 0.044 & 0.009 & 0.198 & 0.055  \\
 50 & 2000 & 0.917 & 0.027 & 0.054 & 0.027 & {\bf 0.212} & 0.059  \\
 50 & 3000 & 0.919 & 0.026 & 0.074 & 0.003 & {\bf 0.212} & 0.058  \\
 50 & 4000 & 0.920 & 0.025 & {\bf 0.084} & 0.001 & 0.203 & 0.053  \\
 75 &   0 & 0.909 & 0.024 & 0.000 & 0.004 & 0.109 & 0.054  \\
 75 & 1000 & 0.913 & 0.024 & 0.000 & 0.000 & 0.203 & 0.054  \\
 75 & 2000 & 0.916 & 0.027 & 0.051 & {\bf 0.033} & 0.207 & {\bf 0.060}  \\
 75 & 3000 & 0.917 & 0.028 & 0.066 & 0.029 & 0.201 & 0.056  \\
 75 & 4000 & 0.918 & 0.027 & 0.075 & 0.020 & 0.207 & 0.059  \\
100 &   0 & 0.909 & 0.024 & 0.000 & 0.004 & 0.109 & 0.056  \\
100 & 1000 & 0.912 & 0.024 & 0.000 & 0.000 & 0.207 & 0.055  \\
100 & 2000 & 0.915 & 0.027 & 0.033 & 0.023 & {\bf 0.212} & 0.059  \\
100 & 3000 & 0.916 & 0.027 & 0.050 & 0.025 & 0.210 & 0.055  \\
100 & 4000 & 0.918 & {\bf 0.028} & 0.065 & {\bf 0.033} & {\bf 0.212} & 0.058
\enddata
\tablecomments{Bold
        quantities indicate the maximum for a given column. Table values 
        represent the average of 3 different instances of 10-fold cross 
        validation.}
\tablenotetext{a}{Percentage of the majority class
        remaining following downsampling.}
\tablenotetext{b}{Percentage increase in the
        minority class via synthetic oversampling (see text).}
\tablenotetext{c}{The
        \recall\ at \precision\ $= 0.1$.}
\tablenotetext{d}{The
        \recall\ at \precision\ $= 0.05$. This is the model FoM.}
\tablenotetext{e}{The
        \precision\ at \recall\ $= 0.1$.}
\end{deluxetable*}

The FoM for the baseline model, which uses the entire training set with no synthetic oversampling, is 0.109. From Table~\ref{tbl:EMP_preds}, it is clear that the synthetic oversampling technique provides a significant improvement over the baseline model, with virtually all oversampled models showing a $\sim$100\% increase in the FoM. Furthermore, it is clear that the precise choice of the degree of over- and downsampling does not have a strong effect on the final model predictions. With the exception of the model featuring 4000\% oversampling and 25\% downsampling, the largest difference in the FoM for oversampled models is $\sim$10\%. Thus, I conclude that the use of a synthetically-oversampled minority class improves the efficiency of EMP star discovery. 

\begin{figure}
\centerline{\includegraphics[width=3.5in]{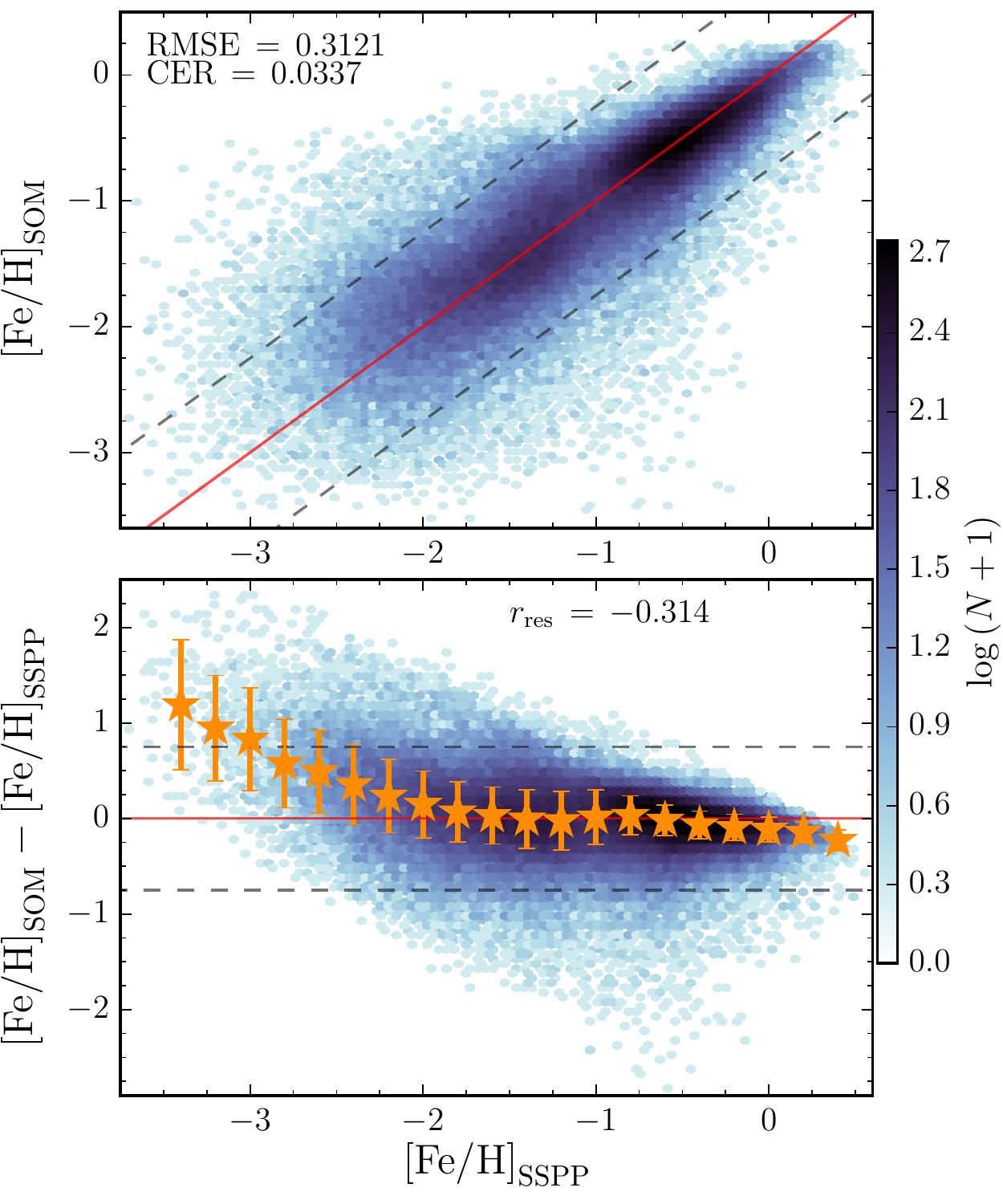}}
\caption[SOM residuals]{10-fold cross-validated results for the full 170,610 star training set using the synthetic-oversampling model with 4000\% oversampling and no downsampling. \textit{Top}: Density plot showing the number of sources in each pixel on the predicted \feh\ vs.\ SSPP \feh\ plane. \textit{Bottom}: Residuals from the model, with the density of sources shown in each pixel. The pixel scale, solid and dashed lines, and orange stars in the top and bottom plots are the same as in Figure~\ref{fig:regress_residuals}. The synthetic-oversampling model produces less-biased predictions, with worse RMSE and CER, than the regression models in \S\ref{sec:results}.}
\label{fig:SOM_residuals}
\end{figure}

Predictions of \feh\ from the synthetic-oversampling model that features 4000\% oversampling and no downsampling are shown in Figure~\ref{fig:SOM_residuals}. Relative to the SVM-regression model, there are far more stars with predicted \feh\ $\le -2.5$ dex. The synthetic-oversampling model is also less biased, as measured by the Pearson $r$ coefficient. The overall performance of the model, as measured by the RMSE, is $\sim$6\% worse than the SVM-regression model and the CER is $\sim$27\% higher than the SVM-regression model. Each model has its relative strengths and weaknesses, and the ultimate choice of model should be driven by a user's science goals. In particular, the synthetic-oversampling model is designed to identify EMP stars, while the SVM-regression model is designed to provide the most accurate estimates of \feh\ for a typical star in the field. Thus, studies focused on metal-poor stars should adopt the synthetic-oversampling model, while studies examining the field should probably adopt the regression model. 

\begin{figure}
\centerline{\includegraphics[width=3.5in]{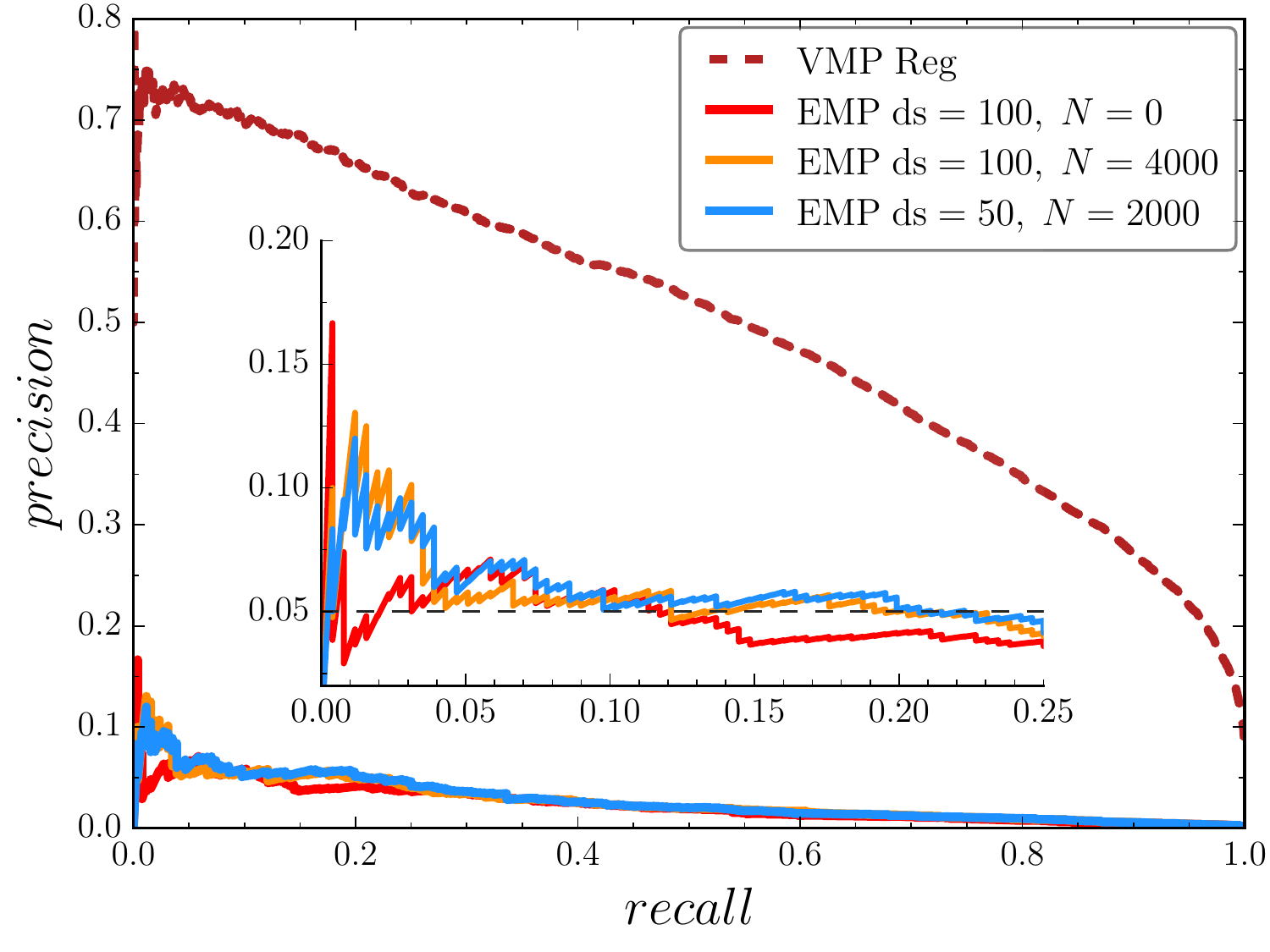}}
\caption[\precision-\recall\ curve]{\precision-\recall\ curves for different parameters of the synthetic oversampling method. The baseline model, which features no oversampling ($N = 0$) or downsampling of the majority ($ds = 100$), is shown with a solid red line. The model with no downsampling and 4000\% ($N = 4000$) oversampling is shown with a solid orange line, while the model with 50\% downsampling ($\mathrm{ds} = 50$) and 2000\% ($N = 2000$) oversampling is shown with the light blue line. The use of synthetic EMP stars significantly improves the performance of the model. The zoom-in shows that the synthetic-oversampling method produces a FoM that is $\sim 2\times$ better than the baseline model. These results are relatively insensitive to the degree of over- and downsampling (see Table~\ref{tbl:EMP_preds}). The maroon-dashed line shows the \precision-\recall\ curve when using the SVM regression model from \S\ref{sec:results} to identify VMP stars.}
\label{fig:PR_curve}
\end{figure}

Example \precision-\recall\ curves are shown in Figure~\ref{fig:PR_curve}. The \precision-\recall\ curves confirm what is shown in Table~\ref{tbl:EMP_preds}: models with synthetic oversampling perform better than the baseline model. In particular, the synthetically-oversampled models show comparable or dramatically improved \precision\ for any \recall\ $\lesssim 0.25$. Interestingly, the synthetic-oversampling method does not provide a significant boost relative to the baseline model for \recall\ $> 0.3$. Figure~\ref{fig:PR_curve} also shows that the differences between the optimized synthetic-oversampling models is small, as was suggested by Table~\ref{tbl:EMP_preds}. It is also worth noting that the majority of false positives for the synthetic-oversampling models are metal poor: $\sim$65\% of the false positives are VMP stars. Finally, note that the SVM regression model presented in \S\ref{sec:results} does a good job of recovering VMP stars without any additional tuning (see the dashed line in Figure~\ref{fig:PR_curve}). The SVM regression model produces a \recall\ $\approx$0.55 at a precision of $\sim$0.5. 

\subsection{Potential Biases in the EMP Sample}\label{sec:EMP_bias}

If the EMP stars in the training set are not representative of the true distribution of EMP stars in the field, then the synthetic-oversampling method will produce a biased sample. It is further possible that synthetic oversampling preferentially selects a specific type of EMP star, such as cool dwarfs or hot sub-giants. If present, these biases would prevent the construction of a complete sample. The number of known EMP stars is small enough, however, that any additional discoveries are valuable for understanding these rare stars. Furthermore, the biases in these methods may be complementary to other methods. For instance, the infrared-color technique presented in \citet{Schlaufman14} preferentially selects giant stars. 

Examining which EMP stars are recovered via CV can provide an estimate of the bias in the synthetic-oversampling method. Figure~\ref{fig:EMP_locations} shows the CV-recovered EMP stars when using a candidate decision threshold of \feh$_\textrm{SOM} <= -2.707$. The results shown are for the model with 4000\% oversampling and no downsampling, though the total number, and location, of sources recovered does not change significantly for any of the models with FoM $\approx 0.2$. From Figure~\ref{fig:EMP_locations}, it is clear that the EMP stars closest to the edges of the stellar locus are the most likely to be recovered. This is not surprising for two reasons: (1) there is higher contrast between these EMP stars and the background of non-EMP stars, and (2) the model has been optimized to sacrifice completeness in favor of \precision. Examination of the SSPP parameters for the recovered EMP stars shows that the synthetic oversampling method is biased towards recovering warm stars with relatively high surface gravity. In particular, of the 126 EMP stars in the training set with \teff\ $\ge 6000 \; \textrm{K}$, $\sim$28\% are recovered, while only $\sim$15\% of the 130 stars with \teff\ $< 6000 \; \textrm{K}$ are recovered. Of the 173 stars with \logg\ $< 3.5 \; \textrm{dex}$, $\sim$14\% are recovered, while $\sim$36\% of the 83 EMP stars with \logg\ $\ge 3.5 \; \textrm{dex}$ are recovered. Thus, it can be concluded that the synthetic oversampling model preferentially selects the hotter, higher surface gravity stars within our training set. 

It is significantly more complicated to determine whether or not the training set is biased relative to the true population of EMP stars. This ir primarily because the actual distribution of EMP stars is unknown, but the complex targeting procedures adopted by the SDSS-I and SDSS-II surveys further muddies the picture. Furthermore, the targeting criteria for the SEGUE portion of SDSS, which is responsible for most of the stellar spectra included in this study, evolved with time to improve the efficiency of target selection \citep{Yanny09}. Spectroscopic targets were selected using a variety of cuts on brightness, photometric color, and proper motion to identify stars belonging to different classes, e.g., white dwarfs, K giants, G stars, etc. As a result, the population of EMP stars detected by SDSS must be biased. In particular, SEGUE used photometric metallicity indicators to preferentially select metal-poor (MP) and metal-poor, turn-off (MPTO) stars. 120 of the 256 EMP stars in the training set were targeted as either MP or MPTO stars. SEGUE biased their search for metal-poor stars toward hotter, and thus more luminous, stars, even though K and M dwarfs live much longer on the main sequence, in order to probe a larger effective survey volume \citep{Yanny09}. As a result, there are few cool EMP stars in the SDSS spectroscopic sample (see the relative lack of EMP stars with $g - r \gtrsim 0.6$ mag in Figure~\ref{fig:EMP_locations}). While a quantitative measure of this bias is not available, it is clear that, by design, the sample of EMP stars identified by SDSS spectroscopy is biased towards F turnoff stars. 

This training set bias provides context for understanding the biased selection of EMP stars from the synthetic-oversampling method. The over-representation of $\sim$F-type stars in the training set is naturally propagated through the machine-learning model to preferentially recover warm (\teff\ $\gtrsim 6000 \; \textrm{K}$) EMP stars.

\subsection{Confirmation of the Synthetic-Oversampling Method with High-Resolution Spectra}

The best way to confirm the efficacy of the synthetic-oversampling method is to obtain spectra of candidate EMP stars, measure \feh\ for these stars, and determine whether or not the \precision\ of the sample is $\approx$0.05, as was predicted in \S\ref{sec:sym_EMP}. Current efforts to obtain such spectra are ongoing and the subject of a future study. In the meantime, the model accuracy can be tested using the high-resolution spectra obtained by \citet{Aoki13}. Using the High Dispersion Spectrograph (HDS) on the Subaru Telescope, \citeauthor{Aoki13} obtained high-resolution spectra of 137 candidate EMP stars selected from the SSPP. Relative to the SSPP, the HDS spectra provide more accurate and precise measurements of \feh, leading to the unambiguous identification of EMP stars. There are 119 stars in common between the \citeauthor{Aoki13} sample and this study. Leave-one-out (LOO) CV\footnote{LOO CV is similar to $k$-fold CV. The difference is that rather than testing on all the data, LOO CV removes a single star from the training set, constructs a model, and then predicts \feh\ for the left-out star. For this study, this procedure is repeated for each of the 119 stars with HDS observations.} is used to measure the fraction of the EMP stars recovered by the synthetic-oversampling method.

\begin{figure}
\centerline{\includegraphics[width=3.6in]{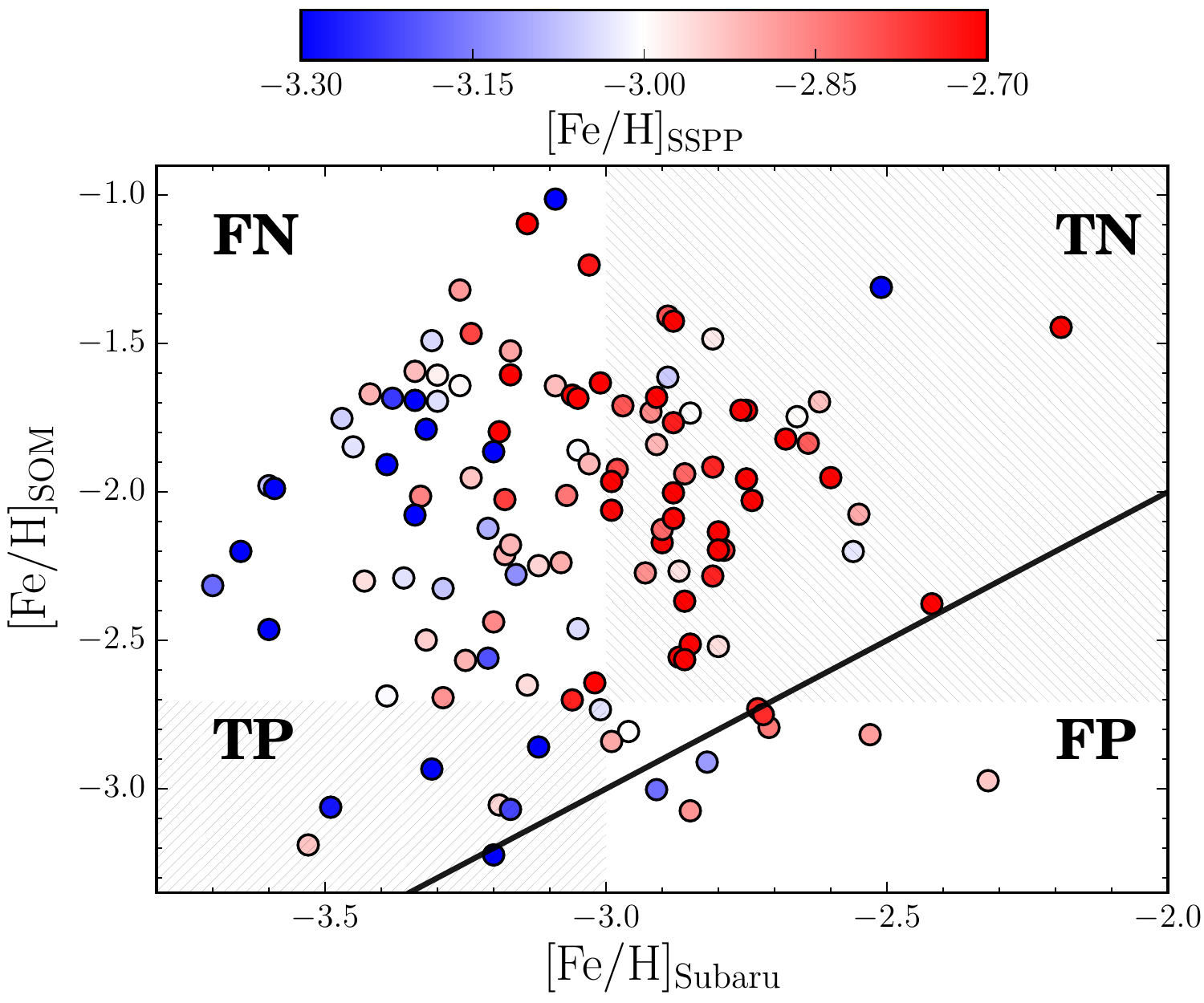}}
\caption[]{Results from the synthetic-oversampling LOO CV performed on the 119 stars in common between this study and the high-resolution spectroscopic sample of \citet{Aoki13}. The synthetic-oversampling method predicted \feh, \feh$_\mathrm{SOM}$, is shown relative to the Subaru/HDS measured \feh. Stars are color coded by their SSPP measured \feh. The solid, black line shows the relation for perfect 1:1 regression. All stars with predicted \feh\ $\le -2.707$ are considered EMP candidates. The areas corresponding to true positives and true negatives are shaded in the lower-left and upper-right corners, respectively. False negatives and false positives are shown in the upper-left and lower-right corners, respectively.}
\label{fig:Subaru_loo}
\end{figure}

Figure~\ref{fig:Subaru_loo} shows the results of the LOO CV procedure, based on a model with 4000\% oversampling and no downsampling. The results show a significant improvement over those shown in Figure~\ref{fig:regress_residuals}, where synthetic oversampling is not employed and there are no stars with predicted \feh\ $\le -2.5$ dex. Of the 119 stars, 64 are genuine EMP stars, and the synthetic-oversampling method identifies 8 of those as EMP candidates based on their photometric colors. This corresponds to \recall\ $= 0.125$, which is worse than expectations (see Table~\ref{tbl:EMP_preds}). More promising is the paucity of false positives, 10, which corresponds to a \precision\ $\approx 0.44$. This estimate of the \precision\ is likely over-optimistic, however, because EMP stars outnumber VMP stars in the \citet{Aoki13} sample. In the halo, VMP stars outnumber EMP stars by a factor of $\sim$50 (e.g., \citealt{Allende-Prieto14}). Pessimistically, this would suggest a ratio of $\sim$500 false positives for every $\sim$8 true positives, corresponding to a \precision\ $\approx$ 0.02. While this is not too dissimilar from the expected \precision\ for the model, the true \precision\ of the model is likely better than 0.02. The \citeauthor{Aoki13} sample of VMP stars is significantly skewed towards stars with \feh\ $\le -2.7$, while the actual halo metallicity distribution strongly favors stars with \feh\ $\approx -2.0$ relative to stars with \feh\ $\le -2.7$. Assuming the model is less likely to identify the most metal-rich VMP stars as EMP candidates, which CV shows is the case, then the \precision\ should be better than the pessimistic estimate. 

Finally, note that the stars shown in Figure~\ref{fig:Subaru_loo} are color coded by \feh$_\mathrm{SSPP}$. While the \recall\ is worse than one would expect based on the CV results from \S\ref{sec:sym_EMP}, it is worth noting that the SSPP slightly over-predicts \feh. In particular, only 36 of the 119 stars have \feh$_\mathrm{SSPP} \le -3.0$, meaning the SSPP sample is biased away from EMP stars, as determined by the Subaru spectra. Relative to the \feh$_\mathrm{SSPP}$ labels, the synthetic-oversampling model produces a \recall\ of 0.25, as one would expect based on the results shown in Table~\ref{tbl:EMP_preds}. Ultimately, this is a demonstration that the results of the machine-learning models are only as good as the training set. When comparing the SSPP measurements to those from the high-resolution spectra, the SSPP has a \recall\ $\approx 0.47$, assuming a class boundary of \feh$_\mathrm{SSPP} = -3.0$. Assuming that the spectroscopic measurements from the Subaru spectra are more accurate than the SSPP, this means the synthetic-oversampling method has a \recall\ ceiling of $\sim$0.47. Moving forward, there are two paths towards improving this ceiling: (i) improve the accuracy of the SSPP measurements, or (ii) obtain significantly more high-resolution spectra, and build a model using those \feh\ measurements. While several incremental improvements have been made to the SSPP (e.g., \citealt{Ahn12,Schlesinger12,Aoki13}), low-resolution spectra will always produce lower-accuracy measurements than their high-resolution counterparts. Furthermore, high-resolution spectra are extremely expensive, meaning a new, uniformly analyzed training set with $> 10^5$ sources is unlikely to be available any time soon.\footnote{SDSS has obtained a set of high-resolution near-infrared spectra that is this large \citep{Alam15}, however, that sample includes very few VMP stars and virtually no EMP stars.} Thus, for the foreseeable future, and despite some clear limitations, the SSPP provides the best basis for a training set to search for EMP stars. 

\section{Final Field-star Predictions}\label{sec:field_preds}

Finally, \feh\ values are predicted for all SDSS stars that satisfy the same selection criteria as the training set (see \S\ref{sec:training_set}). In sum, there are 14,337,770 sources in SDSS DR10 that satisfy all of those photometric criteria, and have \texttt{ProfPSF} = 1, which excludes sources with extended morphologies. Predicted \feh\ values, from both the SVM-regression model (see \S\ref{sec:regress_results}) and the synthetic-oversampling model (see \S\ref{sec:synthetic_oversampling}), for most of these stars are reported in Table~\ref{tbl:field_preds}, though important caveats apply. 

\begin{deluxetable*}{lrcccccc}
\tabletypesize{\small}
\tablecolumns{8}
\tablecaption{Final Metallicity Predictions for Field Stars\label{tbl:field_preds}}
\tablehead{\colhead{Name} & 
    \colhead{Object ID\tablenotemark{a}} & 
    \colhead{$\alpha_\mathrm{J2000.0}$} & 
    \colhead{$\delta_\mathrm{J2000.0}$} & 
    \colhead{\teff\tablenotemark{b}} & 
    \colhead{\feh$_\mathrm{SVM}$\tablenotemark{c}} & 
    \colhead{\feh$_\mathrm{SOM}$\tablenotemark{d}} &
    \colhead{$\rho$\tablenotemark{e}} \\
    \colhead{} & \colhead{} & \colhead{(hh:mm:ss.ss)} & 
    \colhead{(dd:mm:ss.s)} & \colhead{(K)} & \colhead{(dex)} & 
    \colhead{(dex)} & \colhead{} }
\startdata
SDSS J000000.00$+20$4152.5 & 1237680247351279746 & 
                        00:00:00.00 & $+20$:41:52.5 & 
                        5617 & $-2.024$ & 
                        $-2.425$ & 0.0963 \\
SDSS J000000.01$+34$5915.4 & 1237666184574271704 & 
                        00:00:00.01 & $+34$:59:15.4 & 
                        4579 & $-0.568$ & 
                        $-0.567$ & 0.1641 \\
SDSS J000000.02$+12$5954.1 & 1237678920204681228 & 
                        00:00:00.02 & $+12$:59:54.1 & 
                        5903 & $-0.815$ & 
                        $-0.828$ & 0.0627 \\
SDSS J000000.03$+03$2107.2 & 1237678620102164731 & 
                        00:00:00.03 & $+03$:21:07.2 & 
                        5340 & $-0.581$ & 
                        $-0.587$ & 0.0898 \\
SDSS J000000.04$+01$5313.0 & 1237678596479844501 & 
                        00:00:00.04 & $+01$:53:13.0 & 
                        4726 & $-0.307$ & 
                        $-0.311$ & 0.1030 \\
SDSS J000000.05$-00$5019.4 & 1237663783123681350 & 
                        00:00:00.05 & $-00$:50:19.4 & 
                        6110 & $-0.960$ & 
                        $-0.958$ & 0.0466 \\
SDSS J000000.05$+06$5743.2 & 1237669680114106516 & 
                        00:00:00.05 & $+06$:57:43.2 & 
                        5010 & $-0.781$ & 
                        $-0.779$ & 0.0830 \\
SDSS J000000.07$+33$3115.1 & 1237663307989909606 & 
                        00:00:00.07 & $+33$:31:15.1 & 
                        5631 & $-0.394$ & 
                        $-0.393$ & 0.0608 \\
SDSS J000000.08$+20$2502.3 & 1237679504318922768 & 
                        00:00:00.08 & $+20$:25:02.3 & 
                        5467 & \phantom{$-$}$0.166$ & 
                        \phantom{$-$}$0.162$ & 0.1057 \\
SDSS J000000.08$+30$5810.6 & 1237663234451309002 & 
                        00:00:00.08 & $+30$:58:10.6 & 
                        5148 & $-0.356$ & 
                        $-0.348$ & 0.0842
\enddata
\tablecomments{Only the first ten sources are presented 
    here as an example of the form and content of the complete table. 
    The full table, containing all 12,735,277 SDSS point sources with \feh\ 
    predictions, is available online.}
\tablenotetext{a}{\texttt{objID} from the SDSS DR10
        \texttt{PhotoObjAll} table.}
\tablenotetext{b}{Photometrically determined \teff\ using 
        the method of \citet{Pinsonneault12}. See text for further details.}
\tablenotetext{c}{Photometric \feh\ determined using the 
        SVM-regression model from \S\ref{sec:regress_results}.}
\tablenotetext{d}{Photometric \feh\ determined using the 
        synthetic-oversampling method (see 
        \S\ref{sec:synthetic_oversampling}). Note -- stars with 
        \feh$_\mathrm{SOM} \le -2.707$ are EMP candidates.}
\tablenotetext{e}{The proximity measure, $\rho$. See 
        Table~\ref{tbl:prox} for useful thresholds on $\rho$.}
\end{deluxetable*}

The first caveat is that, unlike for the training set, this photometric sample does not have spectroscopic measurements of \teff. Given that the training set only includes stars satisfying $4500 {\rm K} \le T_{\rm eff} \le 7000 {\rm K}$, the machine-learning models will not produce reliable predictions for stars outside this temperature range. To select stars that satisfy this criteria, \teff\ is assigned to the photometric sample using the Color-\teff\ relations in \citet{Pinsonneault12}. These Color-\teff\ relations are calibrated for 
$4080 \; \mathrm{K} \le T_\mathrm{eff} < 7000 \; \mathrm{K}$, which covers the full range of \teff\ included in the training set. As their method is not valid at all temperatures, \citeauthor{Pinsonneault12} caution that the three individual relations are only valid for stars with $0.13 < (g - r)_0 < 1.34$, $0.13 < (g - i)_0 < 1.90$, and $0.07 < (g - z)_0 < 2.21$, respectively. Stars with colors outside this range are excluded from Table~\ref{tbl:field_preds}, which restricts the sample of field stars to 13,004,005. The three Color-\teff\ relations, one each for $(g - r)_0$, $(g - i)_0$, and $(g - z)_0$, are applied to each star and the mean \teff\ is adopted. There are 12,735,277 stars with a mean \teff\ between 4500 K and 7000 K, and they are summarized in Table~\ref{tbl:field_preds}.\footnote{The color-\teff\ relations presented in \citet{Pinsonneault12} are calibrated for dwarf stars at \feh\ $= -0.2$ dex. A change in \feh\ results in a chance in \teff\ at fixed color. As a result the limits placed on the photometrically-determined \teff\ will slightly bias the sample towards metal-poor stars by including some that are cooler than 4500 K and hotter than 7000 K (see Table~3 in \citealt{Pinsonneault12}). Given the magnitude limits on the sample and the rarity of metal-poor stars, the overall contamination is expected to be very small. The color-\teff\ relations have a weak dependence on \logg, with only cool, giants ($T_\mathrm{eff} \lesssim 5000$ K; \logg$ \le 3.5$) requiring corrections. Stars with \logg\ $\approx 2.0$ need the most significant corrections, which, nevertheless, are relatively small ($\Delta T_\mathrm{eff} \lesssim 100$~K). Given the rarity of giants in the SDSS photometric sample (see e.g., \citealt{Ivezic08a}), these corrections are ignored and should not significantly bias the final predictions from the models.} 

The second caveat is that most data-driven methods are not reliable outside the parameter space enclosed by the training set. Figure~\ref{fig:SSPP_summary} shows the training set is confined to a specific location in feature space, i.e.\ the stellar locus. Thus, model predictions for sources within the range of acceptable \teff, but well outside the region defined by the training set, may be unreliable. To aid the user in identifying potentially unreliable estimates of \feh, Table~\ref{tbl:field_preds} includes a proximity measure $\rho$, which measures the relative distance of any given star to the training set. The proximity measure is defined as:
\begin{equation}
    \rho_i = \frac{1}{60}\sum_{j = 1}^{60} \left[\sum_{l = 1}^4(x_{i,l} - x_{j,l})^2\right]^{1/2},
\end{equation}
where $\rho_i$ is the proximity measure of the $i^\mathrm{th}$ source, $x_i$ is the 4-dimensional feature vector, with each feature, $l$, scaled to the standard normal distribution, respectively, and the sum is over each of the $j$ 60-nearest neighbors to the $i^\mathrm{th}$ source as determined by the KNN algorithm. Thus, $\rho$ represents the mean Euclidean distance between a given source and its 60-nearest-training-set neighbors.\footnote{The choice of 60 neighbors is arbitrary, but the relative ranking of the proximity measure does not change significantly for any choice of $k \gg 1$ neighbors. $k = 60$ was adopted to match the optimized KNN model from \S\ref{sec:regress_results}.} Sources with large $\rho$ are likely to have unreliable estimates of \feh.

\begin{deluxetable}{cr}
\tabletypesize{\small}
\tablecolumns{2}
\tablewidth{140pt}
\tablecaption{Proximity Measure Thresholds\label{tbl:prox}}
\tablehead{\colhead{Percentile} & \colhead{$\rho_\mathrm{t}$} }
\startdata
68 & 0.0843 \\
90 & 0.1310 \\
95 & 0.1705 \\
99 & 0.3883 \\
99.5 & 0.5774 \\
99.7 & 0.7737
\enddata
\tablecomments{The threshold, $\rho_\mathrm{t}$, corresponding to the percentage of training set sources with $\rho \le \rho_\mathrm{t}$. }
\end{deluxetable}

\begin{figure} \centerline{\includegraphics[width=3.5in]{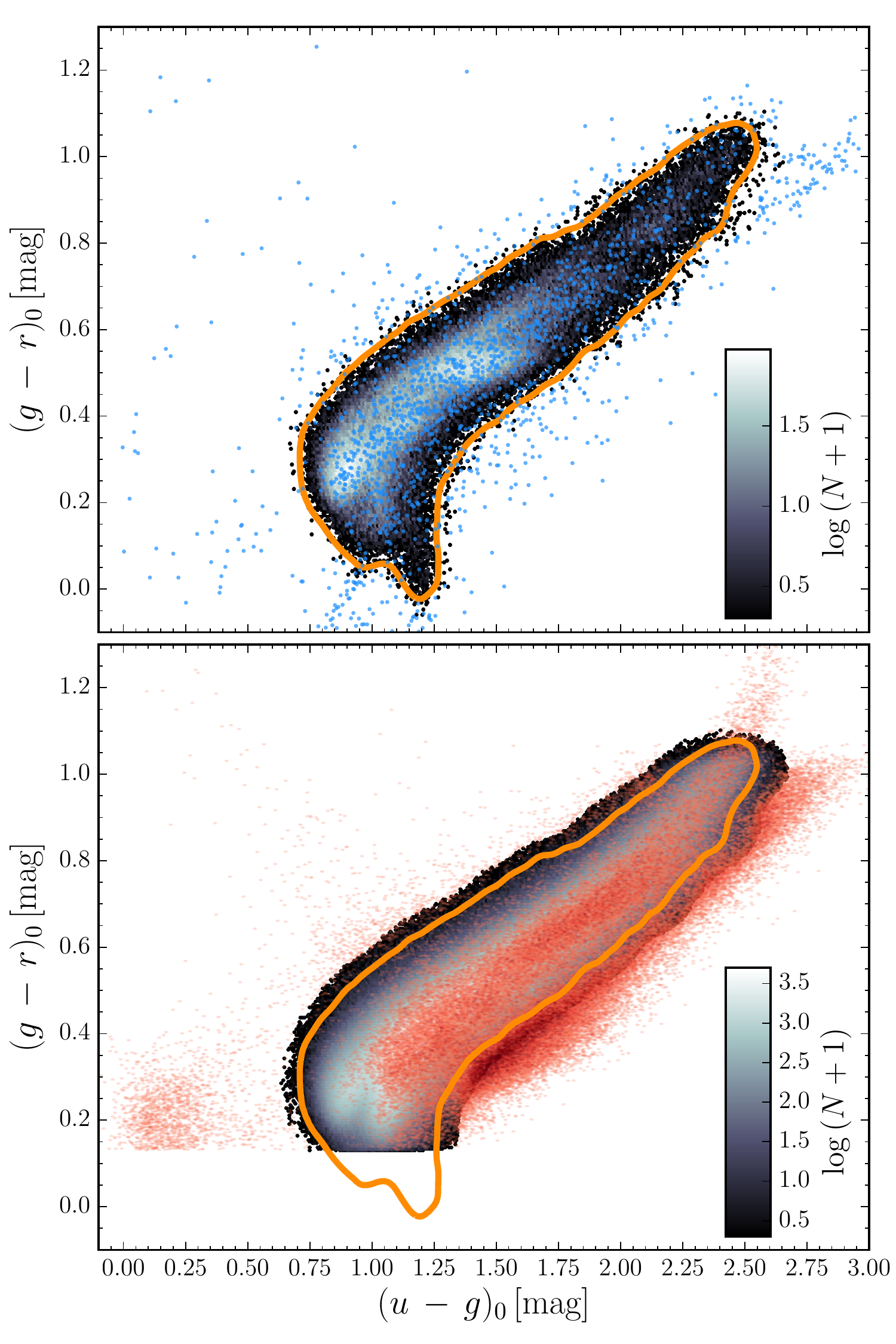}}
    \caption[]{$(u - g)_0$, $(g - r)_0$ CC diagram showing the location of stars with large proximity measure. \textit{Top}: density plot showing the total number of \textit{training set} stars in each $\sim$0.01 $\times$ 0.01 mag pixel on a white to black color scale. The blue points show training set stars in the 99$^\mathrm{th}$ percentile of proximity measure, corresponding to $\rho \ge 0.3883$. The solid orange line shows the 99.7\% contour for the training set, as measured in the $(u - g)_0$, $(g - r)_0$ plane. \textit{Bottom}: density plot showing the total number of \textit{field stars} with $\rho < 0.3883$ per pixel. The solid orange line shows the same contour as the top panel. Red points show the location of field stars with $\rho \ge 0.3883$. The majority of stars outside the main stellar locus have large proximity measure. There are no field stars with $(g - r)_0 \lesssim 0.13$, because the Colors-\teff\ method does not apply to stars with \teff\ $> 7000$ K.}
    \label{fig:proxmeasure}
\end{figure}

The proximity measures are relative, meaning there is no hard and fast rule for a threshold on $\rho$ that eliminates all unreliable \feh\ estimates. Table~\ref{tbl:prox} shows the proximity measure for training set sources based on several commonly adopted threshold percentiles. Studies that require high-fidelity \feh\ estimates can adopt a small $\rho$ threshold, while studies requiring larger samples can relax that criterion. Figure~\ref{fig:proxmeasure} shows training-set and field stars that would be considered unreliable when adopting a proximity-measure threshold of $\rho_\mathrm{t} = 0.3883$, corresponding to the most distant 1\% training-set stars. The top panel of Figure~\ref{fig:proxmeasure}, which highlights sources in the training set, shows that nearly every training-set source outside the 99.7\% $(u - g)_0$, $(g - r)_0$ contour is flagged as unreliable. Sources with $\rho > 0.3883$ inside the contour have anomalous $(g - i)_0$ or $(g - z)_0$ colors. Applying the same threshold to the field stars in Table~\ref{tbl:field_preds}, shown in the bottom panel of Figure~\ref{fig:proxmeasure}, shows that sources distant from the training set are flagged as unreliable. In particular, once again the vast majority of stars outside the 99.7\% contour are flagged as unreliable. It is reassuring that the cluster of sources located at $(u - g)_0$, $(g - r)_0 \approx (0.15, 0.2)$, which should be dominated by quasars \citep{sesar07}, is flagged with large $\rho$. Of the 12,735,277 field stars to which the model is being applied, the $\rho_\mathrm{t} = 0.3883$ threshold would flag $\sim$1.7\% as potentially unreliable. That this number is close to 1\% suggests that the distribution of stars in the training set and the field are very similar. 

The synthetic-oversampling predictions (\feh$_\mathrm{SOM}$) presented in Table~\ref{tbl:field_preds} come from the model with 4000\% oversampling and no downsampling. When using this model, any stars with \feh$_\mathrm{SOM} \le -2.707$ are considered EMP candidates. Adopting this threshold results in 17,605 candidates in Table~\ref{tbl:field_preds}. That threshold corresponds to a \precision\ = 0.05, meaning $\sim$880 of these candidates should be genuine EMP stars. This estimate ignores the proximity measure, however, and thus likely overestimates the true number of EMP stars in the sample. The application of a conservation proximity-measure threshold, $\rho \le 0.1705$, which corresponds to the 95$^\mathrm{th}$ percentile of the training set (see Table~\ref{tbl:prox}), reduces the sample to 11,491,213 stars with 11,849 EMP candidates. Of these candidates, $\sim$590 should be bonafide EMP stars given the \precision\ of the synthetic-oversampling model. There are only a few hundred known EMP stars that have been confirmed with high-resolution spectra (e.g., \citealt{Aoki13,Roederer14,Jacobson15}), the discovery of $\sim$600 new members of the class would represent a huge windfall for this field of study. 

\section{Summary and Conclusions}\label{sec:conclusions}

I have presented a new photometric method for inferring stellar metallicity from the SDSS $ugriz$ filters. The model, which utilizes machine-learning algorithms, is capable of identifying previously unknown EMP stars once the training set has been supplemented with synthetic EMP stars. The model is trained using a large sample of SDSS stars with high SNR spectra and reliable measurements of \feh\ from the SSPP. Following reasonable cuts on photometric and spectroscopic quality, and the removal of duplicate spectra of the same star, the training set consists of 170,610 unique stars.  

The \feh\ regression model represents an improvement over previous methods by utilizing all four non-redundant colors and a non-parametric model capable of capturing complex interactions between the colors. Three separate models, $k$-nearest neighbors, random forest, and support vector machines, are trained, and optimized, using 110,000 stars from the training set. When these models are applied to the 60,610 star independent validation set, they each produce an RMSE $\approx 0.29$ dex relative to the spectroscopic measurements of \feh. Of the three models, SVM produces the smallest RMSE and bias, though the improvement relative to $k$NN and RF is small ($\lesssim$1\%). The performance of the machine-learning models is compared to that of low-resolution spectra, which produce a typical scatter of $\sim$0.24 dex when measuring \feh\ \citep{lee08}. Using a sample of stars with high-resolution spectroscopic observations as ground truth, the SSPP provides only a $\sim$17\% improvement over the photometric method presented in this paper. Thus, the \feh\ regression methods presented here are comparable to the accuracy achieved with low-resolution spectra, with the major benefit that photometric colors can be acquired much cheaper than spectra. Furthermore, it was demonstrated that the machine-learning regression methods perform better than other photometric \feh\ techniques, while also being more general. In particular, there is an $\sim$18\% improvement relative to the methods presented in \citet{Ivezic08a} and \citet{Bond10}. As a demonstration of the fidelity of the model, \feh\ predictions for 12,735,277 stars without spectroscopic observations are presented in Table~\ref{tbl:field_preds}. Proximity measures are provided for the $\sim$12 million stars with \feh\ predictions, in order to evaluate the reliability of the individual estimates. 

A challenge for this method, and all photometric-metallicity techniques, is correcting for interstellar reddening. The ability to measure \feh\ directly from absorption lines, independent of reddening, remains a major advantage of spectroscopy. In principle, data-driven photometric methods could be used to recover \teff, \feh, and extinction (the method presented in \S\ref{sec:results} effectively recovers \teff\ and \feh), but that would require a significantly enhanced training set. Typically, a training set must grow by $\sim$an order of magnitude to properly capture the diversity necessary to resolve a new parameter. Even with such an expanded training set, I speculate that broadband filters will struggle to fully break the degeneracies between these parameters (unless there is also a significant improvement in photometric precision). Thus, broadband photometric-metallicity techniques are, and will remain, limited in the vicinity of the Galactic plane. 

A primary aim of developing the photometric model was to discover EMP stars. There is a significant class imbalance in the training set, $< 0.2$\% of the sample consists of EMP stars, making it difficult to identify these rare relics of the early universe. To improve the recoverability of these sources, a new framework, referred to as the synthetic-oversampling method, was developed where synthetic EMP are added to the training set while a randomly selected fraction of the majority (non-EMP) class stars are removed. 

The goal of the synthetic-oversampling method is to identify EMP stars, while having a relatively low tolerance for false positives. Thus, the adopted FoM is to maximize the model \recall\ at a fixed \precision\ $= 0.05$, which corresponds to 19 false positives for every newly discovered EMP star. It is found that the synthetic-oversampling method outperforms the baseline model, where no oversampling or downsampling have occurred, with a \recall\ $\approx$ 0.2 at \precision\ $= 0.05$. This represents a $\sim$100\% increase in the FoM relative to the baseline model. The synthetic oversampling method was further tested using 119 stars with high-resolution spectroscopic observations from \citet{Aoki13}. This sample includes 64 bonafide EMP stars, and the use of leave-one-out CV shows that the model produces a \recall\ $\approx 0.125$. 

An examination of the EMP stars that are recovered by the synthetic-oversampling method shows that there is a bias towards the selection of warm (\teff\ $\gtrsim 6000 \; \mathrm{K}$) stars with relatively high surface gravities (\logg\ $\gtrsim$ 3.5). The SEGUE target selection of metal-poor stars was intentionally biased towards warmer stars \citep{Yanny09}, which probe larger volumes at fixed mag, which, in turn leads to a bias in the training set for this study. Thus, the bias introduced by SEGUE is propagated through to the synthetic-oversampling method.

While 19 false positives for every EMP star seems high, this represents a significant improvement over the metal-poor candidate selection techniques adopted by SEGUE. In particular, within the training set 20,200 stars were targeted as metal-poor, and only $\sim$0.3\% are EMP stars. Another 18,606 were targeted as likely metal-poor turnoff stars, with, again, a yield of only $\sim$0.3\%. Furthermore, of the 19 false positives for every EMP star, $\sim$65\% of those false positives are VMP stars. Thus, the synthetic-oversampling method produces a highly pure sample of metal-poor stars. Future and ongoing spectroscopic surveys hoping to efficiently identify large samples of EMP stars, such as the Large Sky Area Multi-Object fiber Spectroscopic Telescope (LAMOST; \citealt{Cui12, Deng12}), should adopt the synthetic-oversampling method for target selection.

Astronomy has embarked upon an age where wide-field photometry is cheap: SDSS and Pan-STARRS \citep{Kaiser10} have mapped a significant fraction of the sky in multiple filters to a depth of $\sim$21-22 mag. LSST will do the same for the entire southern sky to a depth of $\sim$27 mag. Now, more than ever, it is imperative that meaningful physical information, such as \feh, can be extracted from photometric-only surveys. The large volume of data produced by LSST will prove no better than existing observations if the proper algorithmic solutions are not developed to deal with the new, complex data stream. Machine-learning methods provide a promising way to cope with the coming data deluge, and this work serves as a step in that direction. A great deal can be learned about the Galaxy from photometric metallicity measurements (e.g., \citealt{Ivezic08a,Bond10}), while the heaps of yet to be discovered EMP stars provide the promise of shedding light on otherwise unobservable aspects of the early universe. 

\vspace{-0.1in}

\acknowledgments 

I thank B.\ Bue and U.\ Rebbapragada for multiple useful conversations on class imbalance and model optimization. I am grateful that J.\ Cohen was willing to suffer many (possibly naive) questions about stellar metallicity measurements and bias in the SDSS sample. J.\ Cohen, L.\ Hillenbrand, and E.\ Kirby provided comments on an early version of this paper, which greatly improved its final content. Finally, I thank the SEGUE team for making the results of the SSPP public, and I am especially indebted to Y.\ S.\ Lee, who has answered many inquiries about the SSPP flags and reliability of the individual \feh\ measurement methods.

I acknowledge support for this work by NASA from a
Hubble Fellowship grant: HST-HF-51325.01, awarded by STScI,
operated by AURA, Inc., for NASA, under contract NAS 5-26555. 
Part of the research was carried out at the Jet Propulsion 
Laboratory, California Institute of Technology, under a contract
with NASA. 

Funding for SDSS-III has been provided by the Alfred P. Sloan Foundation, the Participating Institutions, the National Science Foundation, and the U.S. Department of Energy Office of Science. The SDSS-III web site is http://www.sdss3.org/.

SDSS-III is managed by the Astrophysical Research Consortium for the Participating Institutions of the SDSS-III Collaboration including the University of Arizona, the Brazilian Participation Group, Brookhaven National Laboratory, Carnegie Mellon University, University of Florida, the French Participation Group, the German Participation Group, Harvard University, the Instituto de Astrofisica de Canarias, the Michigan State/Notre Dame/JINA Participation Group, Johns Hopkins University, Lawrence Berkeley National Laboratory, Max Planck Institute for Astrophysics, Max Planck Institute for Extraterrestrial Physics, New Mexico State University, New York University, Ohio State University, Pennsylvania State University, University of Portsmouth, Princeton University, the Spanish Participation Group, University of Tokyo, University of Utah, Vanderbilt University, University of Virginia, University of Washington, and Yale University.

\textit{ Facilities:} 
\facility{Sloan}

\textcopyright 2015. All rights reserved.

\end{document}